\documentclass[secnumarabic,amssymb, nobibnotes, aps, prd]{revtex4-2}
\usepackage{graphicx}
\usepackage{amsmath}
\usepackage{subcaption}
\usepackage{xcolor}
\usepackage{multirow}
\usepackage{natbib}

\begin{document}
	\title{Information-Geometric Signatures from Nonextensivity in the $1$-D Blume-Capel Model}
	\author{Amijit Bhattacharjee$^1$}
	
	\email{$rs_amijitbhattacharjee@dibru.ac.in$}
	\author{Himanshu Bora$^2$}
	
	\email{$hb@kamrupcollege.ac.in$}
	\author{Prabwal Phukon$^{1,3}$}
	\email{prabwal@dibru.ac.in}	
	\affiliation{$1.$Department of Physics, Dibrugarh University, Dibrugarh, Assam,786004.\\$2.$ Department of Physics, Kamrup College, Chamata, Nalbari, Assam - 78306, India\\$3.$Theoretical Physics Division, Centre for Atmospheric Studies, Dibrugarh University, Dibrugarh, Assam,786004.}
	\begin{abstract}

We study the thermodynamic geometry of the one-dimensional Blume--Capel model within the Tsallis nonextensive framework to understand how generalized statistics modify correlation structure and pseudo-critical behaviour. Using the transfer matrix method, we construct the Tsallis entropy based thermodynamic metric as its negative Hessian on the parameter space $(\beta, J)$, with the crystal-field anisotropy $D$ as a control parameter, and compute the associated scalar curvature $R(T)$ as a measure of correlations. Although no true phase transition occurs in one dimension, $R(T)$ exhibits finite peaks signaling pseudo-critical crossovers. We analyze both $D < J$ and $D > J$ regimes and show that deviations from the Boltzmann--Gibbs limit ($q=1$) systematically deform the curvature profile: for $q>1$ the peak shifts and correlations persist beyond the crossover, whereas for $q<1$ the peak is weakened or suppressed. We further compare the scalar curvature with the specific heat capacity and find that the temperature assoiciated with the curvature extrema has significant overlap with the temperatures at which the specific heat produces sudden fluctuations, demonstrating that the geometric signatures consistently identify the pseudo-critical crossovers. Our results demonstrate that the Tsallis parameter $q$ geometrically reshapes the entropy surface, providing a clear information-geometric interpretation of nonextensive effects in spin-1 systems.

\end{abstract}

	\maketitle
	\section{Introduction}

The thermodynamic geometric approach provides a profound and unified framework for describing equilibrium thermodynamics through Riemannian structures defined on the manifold of thermodynamic states. Originally introduced by Frank Weinhold \cite{weinhold1975} and further developed by George Ruppeiner \cite{ruppeiner1979,ruppeiner1995review}, this formalism associates a metric tensor to the space of equilibrium parameters, with the scalar curvature $R$ encoding information about the underlying microscopic correlations. In this geometric picture, a vanishing curvature corresponds to non-interacting systems, while large or divergent curvature signals strong correlations and critical behaviour. Over the years, thermodynamic geometry has emerged as a powerful diagnostic tool in diverse contexts, including magnetic systems \cite{janyszek1989magnetic}, quantum gases \cite{janyszek1990ideal}, black holes \cite{aman2008ruppeiner,soroushfar2016thermodynamic,bhattacharjee2025btz}, and condensed matter or spin-lattice models \cite{sahay2020thermodynamic,sanwari2022thermodynamic,ruppeiner2015frustration}.
\\ 

Among lattice spin systems, the Blume--Capel (BC) model \cite{blume1966theory,capel1966possibility} occupies a distinguished position as a natural generalization of the Ising model. Unlike the spin-$\frac{1}{2}$ Ising system, the Blume--Capel model incorporates an additional spin state $S_i = 0$, representing a non-magnetic or vacancy configuration, alongside the magnetic states $S_i = \pm 1$. The presence of the single-ion anisotropy parameter $D$ introduces a competition between magnetic ordering and vacancy formation, enriching the thermodynamic structure of the model. In higher dimensions, this competition gives rise to tricritical behavior and first-order phase transitions, making the Blume--Capel model a prototypical system for studying multicritical phenomena \cite{saul1974tricritical,gunaratnam2024existence,kwak2015first,moueddene2024critical}. \\ 

Even in one dimension, where no true thermodynamic phase transition occurs at finite temperature, the model exhibits sharp crossovers and pseudo-critical behaviour for specific ranges of the ratio $D/J$ \cite{liu2025pseudo}. These features manifest through pronounced variations in response functions and correlation lengths. From a geometric standpoint, such pseudo-critical effects are reflected in peaks of the thermodynamic scalar curvature $R(T)$, which measures the effective correlation volume \cite{ruppeiner1995review,ruppeiner2010interactions}. The analytical tractability of the one-dimensional case via the transfer matrix method makes it an ideal laboratory for exploring how modifications in statistical structure influence thermodynamic geometry \cite{sahay2020thermodynamic,sanwari2022thermodynamic}. \\

Beyond its role as an analytically tractable theoretical model, the Blume--Capel framework is also part of a broader class of low-dimensional magnetic systems that continue to attract considerable research interest due to their rich critical and quantum properties. Recently, significant attention has been devoted to exploring the rich phase topologies and critical behaviours of low-dimensional magnetic architectures. For instance, the quantum renormalization group approach has been successfully employed to map out quantum phase transitions and scaling behaviors in anisotropic XXZ Heisenberg spin chains \cite{hachem2025quantum}. Concurrently, studies on 2D monolayer configurations, such as the $B_{36}$ borophene-like systems under the Blume-Capel model, have highlighted how transverse crystal fields induce quantum disorder and alter critical transition paths \cite{elabbassi2025phase}. Furthermore, mixed-spin bilayer frameworks like graphene configurations have emerged as prime candidates for nanoscale magnetic devices. Recent Monte Carlo simulations have demonstrated that tuning exchange interactions and crystal field parameters in these systems yields complex, multi-loop hysteresis behaviors highly sensitive to thermal fluctuations \cite{salama2026monte}.\\

Beyond the conventional Boltzmann--Gibbs framework, generalized entropy formalisms have attracted considerable attention in the study of complex systems characterized by long-range interactions, fractal phase-space structures, or memory effects \cite{beck2003superstatistics,gellmann2004nonextensive,abe2001nonextensive,tsallis2009introduction,naudts2011generalised}. In particular, the entropy proposed by Tsallis introduces a deformation parameter $q$ that controls deviations from extensivity \cite{tsallis1988possible}. For $q \to 1$, the standard Boltzmann--Gibbs entropy is recovered, whereas for $q \neq 1$, the statistical weights of microstates are modified, effectively enhancing or suppressing rare configurations depending on the direction of deviation. Such nonextensive frameworks have been successfully applied to gravitational clustering, turbulence, anomalous diffusion, and strongly correlated systems \cite{beck2003superstatistics,tsallis2009introduction}. A brief summary of the Tsallis entropy is as follows:
\\ 

\underline{\textbf{Tsallis Entropy}}:
Put forward by Constantino Tsallis in 1988, Tsallis entropy \cite{tsallis1988possible} is a generalization of the Boltzmann--Gibbs entropy that aims to describe systems exhibiting long-range interactions, memory effects, or multifractal structures under conditions in which the standard additive entropy may fail. It is defined as
\[
S_q = \frac{1 - \sum_i p_i^q}{q - 1},
\]
where \( q \) is the entropic index that measures the degree of non-extensivity. For \( q \to 1 \), Tsallis entropy recovers the usual Boltzmann--Gibbs form. When \( q > 1 \), rare events, i.e. low-probability states, are suppressed, while for \( q < 1 \), they are enhanced. This makes the entropy useful for modelling diverse physical systems such as turbulent flows, gravitational clustering, anomalous diffusion, and strongly correlated systems \cite{beck2003superstatistics,gellmann2004nonextensive,abe2001nonextensive,tsallis2009introduction,naudts2011generalised}. Its non-additive property makes it particularly suitable for complex systems where correlations between subsystems are significant. \\

Here the exponent $q$ modifies the sensitivity of the entropy to various probability values. When $q > 1$, higher probabilities are effectively emphasized because raising $p_i \le 1$ to a power greater than one suppresses smaller probabilities more strongly than larger ones. Consequently, rare events contribute less, and the entropy becomes more sensitive to dominant (main) events. In contrast, when $q < 1$, small probabilities are relatively enhanced since $p_i^q$ with $q < 1$ increases the weight of rare events compared to the standard case. As a result, the entropy becomes more sensitive to the tails of the distribution, favouring fluctuations and uncommon configurations. In the limit $q \to 1$, one recovers the usual Boltzmann--Gibbs entropy as the follows:\\

The Tsallis entropy is given as:
\begin{equation}
S_q = \frac{1 - \sum_i p_i^q}{q - 1}.
\label{tsallisformula}
\end{equation}

We rewrite
\begin{equation}
\sum_i p_i^q 
= \sum_i p_i \, e^{(q-1)\ln p_i}.
\label{sumP}
\end{equation}

For $q \to 1$, we have
\[
(q-1) << 1.
\]

Therefore, using the Taylor expansion,
\begin{equation}
e^{(q-1)\ln p_i}
= 1 + (q-1)\ln p_i.
\end{equation}

Substituting the eqtn. \ref{sumP} into eqtn.\ref{tsallisformula}, we get
\begin{align}
\sum_i p_i^q 
&= \sum_i p_i \left( 1 + (q-1)\ln p_i \right) \\
&= \sum_i p_i + (q-1)\sum_i p_i \ln p_i.
\end{align}

Therefore,
\begin{equation}
S_q 
= \frac{1 - \sum_i \left( p_i + (q-1)p_i \ln p_i \right)}{q-1}.
\end{equation}

Since $\sum_i p_i = 1$, we obtain
\begin{align}
S_q 
&= \frac{1 - 1 - (q-1)\sum_i p_i \ln p_i}{q-1} \\
&= - \sum_i p_i \ln p_i.
\end{align}

Hence,
\begin{equation}
\lim_{q \to 1} S_q = S_{BG} = - \sum_i p_i \ln p_i.
\end{equation}

Therefore in the limit $q \to 1$, the Tsallis entropy $S_q$ reduces to the usual Boltzmann-Gibbs entropy as follows:
\[
S = S_{BG} = -\sum_i p_i \ln p_i,
\]

When generalized entropies are employed, thermodynamic potentials and response functions acquire modified functional forms, and consequently the geometry of the entropy surface is altered. Since the thermodynamic metric is defined as the negative Hessian of the entropy with respect to its parameters, any deformation of the entropy directly reshapes the underlying geometric manifold. The scalar curvature therefore serves as a sensitive probe of how nonextensivity modifies correlation strength, stability properties, and pseudo-critical structure.\\

The use of Tsallis entropy in the present work should not be interpreted as implying that the one-dimensional Blume--Capel model intrinsically requires a nonextensive statistical description. Indeed, the conventional model is a short-range interacting system that is exactly solvable within the Boltzmann--Gibbs framework. Rather, our objective is to employ the Blume--Capel model as a controlled and analytically tractable setting in which the consequences of generalized entropy on thermodynamic geometry can be systematically investigated. Since the thermodynamic metric is constructed directly from the entropy through its Hessian structure, replacing the Boltzmann--Gibbs entropy by the Tsallis entropy naturally deforms the underlying thermodynamic manifold. The resulting changes in the scalar curvature therefore provide a direct geometric measure of how nonextensive statistical weighting modifies the correlation structure encoded in the entropy surface.\\

From a broader perspective, Tsallis statistics has been successfully employed as an effective framework for describing systems that exhibit departures from the assumptions underlying conventional extensive thermodynamics, including long-range interactions, memory effects, multifractal structures, and non-ergodic dynamics. In the present study, the nonextensive parameter $q$ is treated as an effective deformation parameter that quantifies deviations from the Boltzmann--Gibbs limit rather than as a fixed material property of a specific experimental realization. The purpose of introducing $q$ is therefore to explore how generalized statistical weighting influences geometric indicators of correlations and pseudo-critical behaviour. The one-dimensional Blume--Capel model, owing to its exact solvability and well-understood thermodynamic properties, provides an ideal benchmark system for isolating and understanding these purely entropy-induced geometric effects. Within this framework, the present work aims to understand how the Tsallis deformation parameter reshapes the thermodynamic state-space geometry of the one-dimensional Blume--Capel model and modifies the associated scalar-curvature signatures of pseudo-critical crossover behaviour. \\

We here analyze the model using the Transfer matrix method which is a well established technique and has been used in many cases. For instance, the transfer-matrix method has been effectively utilized to derive exact expressions for free energy, entropy, and internal energy in one-dimensional lattice architectures, such as the 1D mixed-type Potts-SOS model \cite{akin2024qualitative}, where stability analyses reveal an inherent decay of magnetization to zero and a lack of phase transitions. Conversely, when moving to hierarchical or fractal networks like the semi-infinite Cayley tree \cite{akin2024investigation}, extracting statistical and thermodynamic profiles (e.g., in mixed-spin $(2, 1/2)$ Ising and Blume-Capel systems) often requires a combination of recursive methods, cavity approaches, and transfer-matrix formulations. While tree-like structures frequently demonstrate rich phase transition phenomena in both ferromagnetic and antiferromagnetic domains due to their high coordination connectivity, true 1D chains evaluated strictly via transfer matrices exhibit distinct, continuous thermodynamic behaviors without phase boundaries. Incorporating these dual perspectives allows for a more robust validation of our model's low-dimensional constraints

The objective of the present work is not merely to revisit the thermodynamic geometry of the one-dimensional Blume--Capel model, but rather to investigate how the Tsallis nonextensive entropy deforms the thermodynamic state-space
geometry of the system. In particular, we analyze how the nonextensive parameter $q$ modifies the sign, magnitude, persistence, and temperature location of the thermodynamic scalar curvature. We further examine these effects in both the exchange-dominated ($D<J$) and crystal-field-dominated
($D>J$) regimes and investigate their dependence on system size. To the best of our knowledge, a systematic thermodynamic geometric analysis of the one-dimensional Blume--Capel model within the Tsallis framework has not been reported previously. The present work has its novelty in extending the information-geometric analysis of the $1$-D Blume-Capel model to the Tsallis nonextensive statistical framework and our contribution primarily consists of the following aspects:\\

\begin{enumerate}
   \item \textbf{Introduction of a nonextensive thermodynamic geometry:} 

  We construct the thermodynamic metric from the Hessian of the Tsallis entropy rather than the conventional Boltzmann--Gibbs entropy. This produces a deformed thermodynamic manifold whose geometric properties depend explicitly on the nonextensive parameter $q$.

    \item \textbf{Investigation of how nonextensivity modifies geometric correlations:}

    Rather than merely reproducing known geometric features of the Blume--Capel model, we study how the Tsallis parameter $q$ alters the scalar curvature profile $R(T)$, including changes in the location, magnitude, sign, and persistence of curvature peaks. These modifications provide new insight into how generalized statistical weighting influences thermodynamic correlations.

    \item \textbf{Information-geometric interpretation of the Tsallis deformation:}

    Our analysis demonstrates that the nonextensive parameter acts as a geometric deformation parameter that reshapes the entropy surface and consequently modifies the underlying correlation structure encoded in the scalar curvature. This provides a direct geometric characterization of nonextensive effects in a spin-1 lattice system.

    \item \textbf{Comparison of magnetically dominated and crystal-field dominated regimes:}

    By separately analyzing the cases $D<J$ and $D>J$, we show that the influence of nonextensivity depends strongly on the dominant microscopic configurations of the system. The resulting curvature behaviour reveals distinct geometric responses that are absent in the conventional $q=1$ description.

    \item \textbf{Finite-size geometric analysis within a generalized entropy framework:}

   We further examine how the curvature profiles evolve with system size under Tsallis statistics, providing insight into the interplay between finite-size effects and nonextensive statistical weighting.
\end{enumerate}
\
In the present work, we investigate the one-dimensional Blume--Capel model within the Tsallis nonextensive framework using an entropy-based thermodynamic metric defined on the parameter space $(\beta, J)$, with the crystal field $D$ acting as a control parameter. By constructing the metric tensor
\[
g_{ij} = -\partial_i \partial_j S_q(\beta, J),
\]
and computing the associated scalar curvature $R(T)$, we analyze how the deformation parameter $q$ modifies the curvature--temperature profile and the location and magnitude of pseudo-critical peaks. In addition to the geometric analysis, we compare the thermodynamic scalar curvature with the specific heat capacities calculated within the same Tsallis entropy framework for the $1$-D Blume--Capel model in order to establish that the curvature extrema correspond to physically meaningful pseudo-critical crossovers in the information geometric measure. Our goal is to provide a geometric interpretation of nonextensive effects in spin-1 systems and to clarify how the interplay between exchange interaction and single-ion anisotropy is reshaped under generalized statistics.\\

The remainder of the paper is organized as follows. In Section~\ref{sec:Ising}, we present the 1D Blume-Capel model and transform it with the Tsallis entropy framework. Section~\ref{sec:metric} develops the metric structure and thus its Tsallis generalization. Section~\ref{sec:thermodynamic} is devoted to the thermodynamic geometric analysis of the non-extensive 1D Blume-Capel model. We provide a discussion on the obtained results in Section \ref{sec:discussion}. We conclude with a short summary in Section~\ref{sec:conclusion}.

\section{The One-Dimensional Blume Capel Model}
\label{sec:Ising}

The \textit{Theory of the First-Order Magnetic Phase Change in UO$_2$} \cite{blume1966theory} by Melvin Blume marked the conceptual birth of what later came to be known as the Blume-Capel model. It arose from the need to explain the experimentally observed first-order anti-ferromagnetic transition in UO$_2$, which could not be satisfactorily accounted for by conventional Heisenberg models containing only bilinear exchange interactions. Blume proposed a simple yet profound mechanism: the uranium ion possesses a non-magnetic singlet ground state and a low-lying magnetic triplet split by crystal-field effects, and the exchange interaction treated within a molecular-field approximation, can self-consistently drive a crossing between these levels. As the temperature increases, thermal population of the non-magnetic singlet reduces the exchange splitting, potentially causing a discontinuous collapse of magnetization, \textit{i.e.}, a first-order phase transition. Abstracting from the specific ionic structure of UO$_2$, this physical picture was later reformulated in terms of spin-1 variables with three possible states $(-1, 0, +1)$, where the crystal-field splitting acts as a single-ion anisotropy term that energetically favours or disfavours the $S = 0$ state. The resulting minimal Hamiltonian consisting of bilinear exchange together with a crystal-field term became the Blume-Capel model, providing a unified framework for the study of tri-criticality, first and second-order phase transitions, and vacancy-like spin states in magnetic systems.\\

As is described above, the Blume--Capel (BC) model is a spin--1 generalization of the Ising model, originally introduced to describe systems with both magnetic and non-magnetic states. Each spin variable $S_i$ can take three possible values,
\[
S_i \in \{-1,\,0,\,+1\},
\]
where the state $S_i = 0$ represents a vacancy or a non-magnetic configuration. 
The Blume-Capel model, originally put forward independently by Blume \cite{blume1966theory} and Capel \cite{capel1966possibility}, has become an important an a foundational framework for studying tri-critical behaviour \cite{saul1974tricritical, gunaratnam2024existence, kwak2015first, goldman1973tricritical, moueddene2024critical, azhari2020tricritical, blote2019revisiting, mandal2016geometrical}. One of its strikingly distinguishing features is its capacity to model transitions from second-order to first-order phase transitions and thereby revealing intricate critical behaviours. The model has been investigated using diverse approaches which includes the mean-field theory \cite{blume1966theory, capel1966possibility}, renormalization group techniques \cite{antenucci2014critical}, and Monte Carlo simulations \cite{mahan1978blume, clusel2008alternative, silva2006wang}. Research on the Blume-Capel model has covered a wide range of materials which includes metallic alloys \cite{dias2011site, pena2009blume, blatter1985reversible, sinkler1997neutron}, magnetic thin films \cite{mazzitello2015far}, and superconducting thin films \cite{goldman1973tricritical}. In addition to its applications in physics, the model has been extensively employed in sociology, particularly in recent studies on depolarization \cite{kaufman2024social, diep2024monte}. Owing to the fact that its critical importance and capacity to capture a wide range of phase behaviours the Blume-Capel model has generated a wide range of publications \cite{mozolenko2024blume, cirillo2024homogeneous, akin2024investigation, ovchinnikov2024influence}.

\subsection{Hamiltonian without External Magnetic Field}

In the absence of an external magnetic field, the Hamiltonian of the Blume--Capel model is given by \cite{liu2025pseudo}
\[
\mathcal{H}
=
- J \sum_{\langle i,j \rangle} S_i S_j
+ D \sum_i S_i^2 ,
\]
where:
\begin{itemize}
\item $J$ is the nearest-neighbor exchange interaction strength,
\item $D$ is the single-ion crystal field anisotropy,
\item $\langle i,j \rangle$ denotes a sum over nearest-neighbor pairs.
\end{itemize}

The exchange interaction \(J\) and crystal-field anisotropy \(D\) are the fundamental parameters of the Blume--Capel model. The parameter \(J\) determines the strength of nearest-neighbour magnetic interactions, while \(D\) controls the energetic preference of the non-magnetic \(S=0\) state relative to the magnetic \(S=\pm1\) states. Throughout this work, we set \(J=1\), thereby fixing the energy scale of the system and expressing all thermodynamic quantities in dimensionless units. \\

The first term favors ferromagnetic alignment for $J>0$, while the second term controls the population of the $S_i=0$ state. For large positive $D$, the non-magnetic state is energetically favored, whereas for large $J$, the magnetic states $S_i=\pm1$ dominate.

\subsection{Transfer Matrix Formulation}

For a one-dimensional lattice with periodic boundary conditions, the partition function can be computed exactly using the transfer matrix method. The partition function is
\[
Z = \sum_{\{S_i\}} \exp\left[
\beta J \sum_{i} S_i S_{i+1}
- \beta D \sum_i S_i^2
\right],
\]
where $\beta = 1/(k_B T)$. The transfer matrix $\mathbf{T}$ is defined by
\[
\mathbf{T}_{S,S'}
=
\exp\left[
\beta J S S'
- \frac{\beta D}{2} (S^2 + S'^2)
\right],
\]
with $S, S' \in \{-1,0,+1\}$. Explicitly, in the basis $\{+1,\,0,\,-1\}$, the transfer matrix takes the form
\[
\mathbf{T}
=
\begin{pmatrix}
e^{\beta J - \beta D} & e^{-\beta D/2} & e^{-\beta J - \beta D} \\
e^{-\beta D/2} & 1 & e^{-\beta D/2} \\
e^{-\beta J - \beta D} & e^{-\beta D/2} & e^{\beta J - \beta D}
\end{pmatrix}.
\]

The partition function for a chain of $N$ spins is then
\[
Z = \mathrm{Tr}\, \mathbf{T}^N,
\]

\[
Z(\beta, J, D) = E_1(\beta,J,D)^N + E_2(\beta,J,D)^N + E_3(\beta,J,D)^N
\]

where $E_1(\beta, J, D)$, $E_2(\beta, J, D)$ and $E_3(\beta, J, D)$ are the eigenvalues of the transfer matrix $T$. They are given as

\[
E_1(\beta, J, D) = 2 e^{-\beta D} \sinh(\beta J)
\]

\[
E_2(\beta, J, D) =
\frac{1}{2}
\left[
1 + 2 e^{-\beta D} \cosh(\beta J)
+
\sqrt{\left(2 e^{-\beta D} \cosh(\beta J)-1\right)^2
+ 8 e^{-\beta D}}
\right]
\]

\[
E_3(\beta, J, D) =
\frac{1}{2}
\left[
1 + 2 e^{-\beta D} \cosh(\beta J)
-
\sqrt{\left(2 e^{-\beta D} \cosh(\beta J)-1\right)^2
+ 8 e^{-\beta D}}
\right]
\]

To isolate the extensive bulk behavior from finite-size effects, we must factor out the strictly dominant eigenvalue. As we shall later see, this also helps in computational stability of the numerical   calculations. \\

We can argue that $E_2$ is always the largest eigenvalue for any physical parameter set ($\beta (T) , J, D$ being real and finite) invoking the Perron-Frobenius Theorem, which states that for any matrix with strictly positive entries, there is a unique largest eigenvalue and its corresponding eigenvector is the only one whose components can be chosen to be all strictly positive. The transfer matrix $T$ consists entirely of strictly positive exponential terms ($T_{ij} > 0$ for all physical parameters. The eigenvalue $E_1$ corresponds to the antisymmetric eigenvector $(1, 0, -1)^T$ and since this eigenvector contains a negative component, so $E_1$ can not be the dominant eigenvalue. We are therefore left with the roots $E_2$ and $E_3$ from the symmetric subspace.  $E_2$ is formed by adding the strictly positive principal square root of the discriminant, and hence $E_2 > E_3$ is unconditionally true. Therefore, $E_2$ must be the unique dominant eigenvalue. \\

Factoring out the dominant eigenvalue $E_2$, the partition function is rewritten as:
\begin{equation}
Z(\beta, J, D) = E_2(\beta)^N \left( 1 + \left(\frac{E_1(\beta)}{E_2(\beta)}\right)^N + \left(\frac{E_3(\beta)}{E_2(\beta)}\right)^N \right) = E_2(\beta)^N \big(1 + \Omega_N(\beta)\big)
\end{equation}
where $\Omega_N(\beta) = (E_1/E_2)^N + (E_3/E_2)^N$ acts as the finite-size correction term. As $N \to \infty$, the term $\Omega_N \to 0$ exponentially.

The Tsallis entropy is defined as:
\begin{equation}
S_q = \frac{1 - \sum_i p_i^q}{q - 1},
\label{eq:Tsallis_def}
\end{equation}
which can be rewritten in terms of the partition function using the canonical probabilities,
\begin{equation}
S_q = \frac{1}{q-1} \left( 1 - \frac{Z(q\beta)}{Z(\beta)^q} \right)
\end{equation} \\
Defining the bulk deformation potential $\Delta_q \equiv q \ln E_2(\beta) - \ln E_2(q\beta)$, we substitute the factored partition function to obtain a numerically stable formulation:
\begin{equation}
S_q = \underbrace{\frac{1 - e^{-N \Delta_q}}{q-1}}_{S_{\text{bulk}}} + \underbrace{\frac{e^{-N \Delta_q}}{q-1} \left[ 1 - \frac{1+\Omega_N(q\beta)}{(1+\Omega_N(\beta))^q} \right]}_{S_{\text{corr}}}
\label{eq:tsallis_numerical}
\end{equation}

\begin{figure}[h!]
    \centering
    \includegraphics[width=0.99\textwidth]{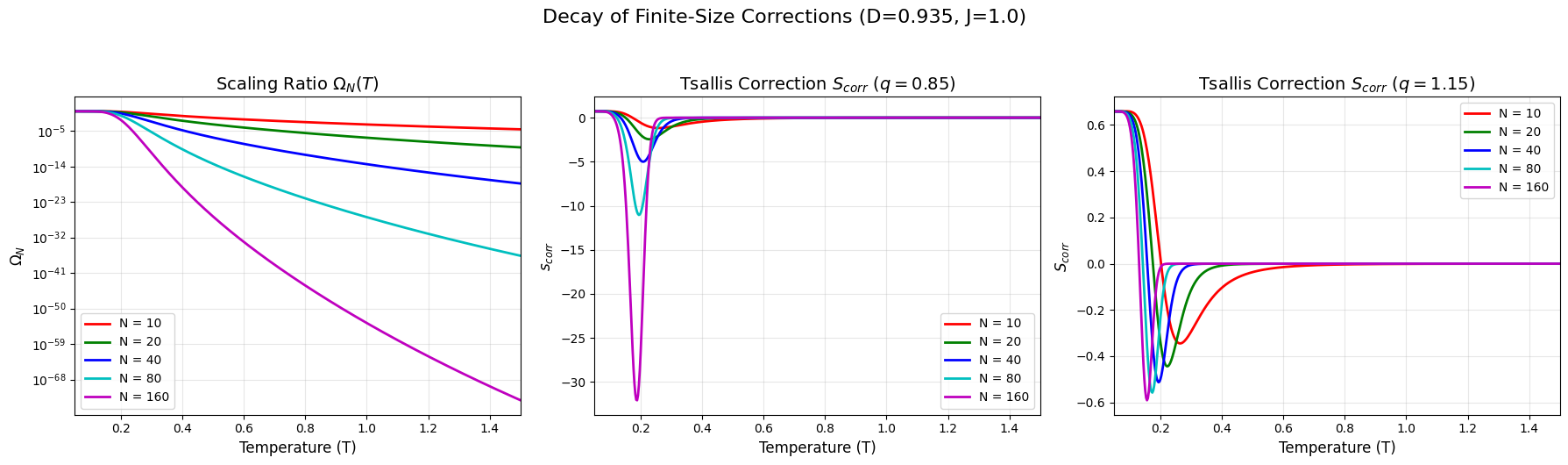}
    \caption{ Scaling of the finite-size correction term $\Omega_N$ as a function of temperature for various system sizes $N$ (left); decay of $S_{corr}$ in $q<1$ (centre) and $q>1$ (right) regimes as a function of temperature for various system sizes $N$. As $N$ increases, the finite-size correction term $\Omega_N$ vanish rapidly and $S_{corr}$ approaches 0, isolating the bulk geometry. We have considered $D=0.935$, $J=1.0$ }
    \label{fig:scaling}
\end{figure}

To illustrate the suppression of the finite-size corrections as $N$ increases, Figure \ref{fig:scaling} demonstrates the decay of the correlation ratio $\Omega_N$ and $S_{corr}$.

\subsection{Remarks}

Although the one-dimensional Blume-Capel model does not exhibit a true phase transition at finite temperature, it shows sharp crossovers or pseudo-critical behaviour for certain values of $D/J$. These features become particularly relevant when generalized entropy formalisms, such as Tsallis entropy, is employed, where effective long-range correlations can emerge even in low-dimensional systems.\\

In addition to that there is another point we would like to discuss regarding the use of Tsallis statistics here. In particular, we would like to state that there are two different frameworks that one can apply under the realm of Tsallis statistics.\\

The first framework corresponds to a fully self-consistent nonextensive thermostatistics, in which the equilibrium probability distribution itself is modified and is obtained by maximizing the Tsallis entropy under appropriate constraints. This leads to the well-known $q$-exponential canonical distribution together with escort probabilities and generalized expectation values.\\

The second framework, which is the one adopted in the present work, retains the conventional Boltzmann--Gibbs canonical probability distribution and evaluates a generalized entropy functional on that equilibrium ensemble. In our analysis, we begin with the ordinary canonical probabilities
\[
p_i=\frac{e^{-\beta E_i}}{Z(\beta)},
\qquad
Z(\beta)=\sum_i e^{-\beta E_i},
\]
which satisfy the standard normalization condition
\[
\sum_i p_i=1.
\]
The Tsallis entropy is then introduced as the generalized information measure
\[
S_q=\frac{1-\sum_i p_i^q}{q-1}.
\]
Substituting the canonical probabilities into this definition immediately gives
\[
\sum_i p_i^q
=
\frac{1}{Z(\beta)^q}
\sum_i e^{-q\beta E_i}
=
\frac{Z(q\beta)}{Z(\beta)^q},
\]
leading to the expression used throughout the manuscript,
\[
S_q=
\frac{1}{q-1}
\left(
1-\frac{Z(q\beta)}{Z(\beta)^q}
\right).
\]
We would like to emphasize that our objective is not to formulate a complete nonextensive thermostatistics based on generalized canonical distributions. Rather, the purpose of this work is to investigate how the use of the Tsallis entropy \emph{functional itself} modifies the information-geometric structure associated with the equilibrium states of the one-dimensional Blume--Capel model. In this sense, the deformation is introduced at the level of the entropy functional, while the underlying equilibrium probability measure remains the standard canonical one.\\

The advantage of this construction is that it allows us to isolate the purely information-theoretic effects of the generalized entropy from the additional modifications that would arise from replacing the underlying canonical distribution by a fully nonextensive $q$-canonical ensemble. Consequently, the resulting changes in the thermodynamic metric and scalar curvature can be directly attributed to the nonadditive nature of the entropy functional itself.

\section{Thermodynamic Metric and Geometric Structure}
\label{sec:metric}
\subsection{Motivation and Conceptual Basis}

Thermodynamic geometry offers a powerful formalism that connects macroscopic thermodynamic behavior with the differential geometry of the entropy or energy surfaces. The idea originated from the works of Weinhold\cite{weinhold1975} and Ruppeiner \cite{ruppeiner1979,ruppeiner1995review}, who independently proposed Riemannian metrics on the thermodynamic state space. In this approach, the equilibrium manifold is endowed with a metric whose curvature encodes the strength and nature of thermodynamic correlations among microscopic degrees of freedom. The scalar curvature \(R\) derived from this metric provides a geometric interpretation of phase transitions, criticality, and statistical interactions. For instance, \(R=0\) corresponds to an ideal gas where microstates are uncorrelated, while a diverging \(R\) signifies critical behaviour where correlation length diverges \cite{janyszek1989magnetic, janyszek1990stability, ruppeiner2010interactions}.\\

The construction of the thermodynamic metric is grounded in fluctuation theory. Consider the probability distribution of fluctuations around equilibrium in an ensemble characterized by extensive variables \(x^i\) and their conjugate intensive variables \(y_i\). In the Gaussian approximation, the probability of a fluctuation \(dx^i\) away from equilibrium is given by
\begin{equation}
P \propto \exp\!\left( -\frac{1}{2} g_{ij}\, dx^i dx^j \right),
\label{eq:fluctuation_prob}
\end{equation}
where \(g_{ij}\) acts as the inverse covariance matrix of the fluctuations and hence defines the thermodynamic metric tensor. For the entropy representation, the line element is defined as
\begin{equation}
ds^2 = - \frac{\partial^2 S}{\partial x^i \partial x^j} \, dx^i dx^j,
\label{eq:ruppeiner_metric}
\end{equation}
which ensures that the entropy is maximized at equilibrium, implying that the metric is positive definite. This formulation is equivalent to defining the metric as the Hessian of the Massieu potential, linking it directly to the second derivatives of the entropy with respect to its natural variables. Weinhold’s earlier definition, in contrast, was based on the internal energy representation,
\begin{equation}
g_{ij}^{(W)} = \frac{\partial^2 U}{\partial x^i \partial x^j},
\end{equation}
and it can be shown that the Ruppeiner and Weinhold metrics are conformally related by the temperature:
\begin{equation}
g_{ij}^{(R)} = \frac{1}{T} g_{ij}^{(W)}.
\label{eq:conformal}
\end{equation}
This relation underscores that both metrics convey the same geometric information but differ in their thermodynamic potentials.

\subsection{Geometric Interpretation of the Scalar Curvature}

Using the generalized entropy function \(S(\beta, J)\) corresponding to the Tsallis form, the entropy-based thermodynamic metric on the parameter manifold \((\beta, J)\) is defined as
\begin{equation}
g_{ij} = -\frac{\partial^2 S(\beta, J)}{\partial x^i \partial x^j}, \qquad x^i = (\beta, J).
\label{eq:metric}
\end{equation}
The associated Ricci scalar curvature \(R(\beta, J)\) is computed from the metric tensor, and its temperature dependence \(R(T)\) provides a geometric measure of thermodynamic correlations. In the nonextensive regime, deviations in \(R(T)\) relative to the Boltzmann–Gibbs case reflect how the generalized entropy modifies the effective correlation volume and pseudo-critical structure of the spin chain.\\

In non-extensive statistical mechanics, the entropy no longer scales additively with subsystem size due to the presence of long-range correlations or fractal phase-space structures. Consequently, the fluctuation geometry must be adapted to incorporate the deformation introduced by generalized entropies such as those of Rényi and Tsallis. The underlying principle remains the same: the metric measures the local curvature of the entropy surface in the space of control parameters, reflecting how stable the equilibrium state is under small perturbations.\\

Given a generalized entropy \(S(\beta,J)\), depending on intensive variables such as the inverse temperature \(\beta\) and coupling constant \(J\), we define the metric as
\begin{equation}
g_{ij} = - \frac{\partial^2 S(\beta,J)}{\partial x^i \partial x^j}, \qquad x^i = (\beta, J),
\label{eq:metric_generalized}
\end{equation}
where the negative sign ensures positive definiteness of the metric because entropy is a concave function of its natural variables. This formulation departs from the traditional extensive representation where the coordinates are extensive variables like \(U\), \(V\), or \(N\). Instead, it constructs an \emph{entropy-based metric on the parameter manifold}, an approach that is particularly useful for systems such as spin chains or black holes, where \(\beta\) and \(J\) are more natural variables than energy or volume \cite{mirza2007ruppeiner, aman2003geometry}.\\

The physical meaning of the components of this metric can be interpreted as follows:
\begin{align}
g_{\beta\beta} &= -\frac{\partial^2 S}{\partial \beta^2} \;\; \Rightarrow \;\; \text{fluctuations in temperature}, \nonumber\\
g_{\beta J} &= -\frac{\partial^2 S}{\partial \beta \partial J} \;\; \Rightarrow \;\; \text{cross-correlation between temperature and coupling}, \nonumber\\
g_{JJ} &= -\frac{\partial^2 S}{\partial J^2} \;\; \Rightarrow \;\; \text{fluctuations in the interaction strength}.
\label{eq:metric_components}
\end{align}
The determinant of the metric \(\det(g_{ij})\) quantifies the joint stability of the system with respect to simultaneous changes in \(\beta\) and \(J\), and its inverse defines the volume element in parameter space associated with thermodynamic fluctuations.\\

In the present work, the thermodynamic state space is constructed using the coordinates \(x^{i}=(\beta,J)\), where \(\beta\) governs thermal fluctuations and \(J\) characterizes the strength of the magnetic exchange interaction. These variables constitute the principal thermodynamic control parameters governing the equilibrium behaviour of the system and therefore provide a natural two-dimensional manifold for investigating the information-geometric structure. The crystal-field anisotropy \(D\) is treated as an external control parameter that is held fixed while the thermodynamic geometry is examined for different physical regimes. This approach allows us to isolate the influence of thermal and interaction-induced fluctuations on the scalar curvature while systematically comparing the exchange-dominated (\(D<J\)) and crystal-field-dominated (\(D>J\)) regimes. \\

We do not claim here that the metric we use here is the standard Ruppeiner metric expressed in the intensive coordinates. Rather, the construction adopted here belongs to the broader class of entropy-Hessian parameter-space metrics, where the generalized entropy is regarded as a scalar function defined on the manifold of external control parameters. In the present work, the natural control parameters of the one-dimensional Blume--Capel model are the inverse temperature $\beta$ and the exchange coupling $J$, while the crystal-field anisotropy $D$ is treated as a fixed control parameter. Consequently, the metric is defined on the two-dimensional parameter manifold
\[
x^i=(\beta,J),
\]
through the Hessian
\[
g_{ij}
=
-
\frac{\partial^2 S_q(\beta,J)}
{\partial x^i \partial x^j}.
\] 

The reason for this choice originates from equilibrium fluctuation theory. In the Gaussian approximation, the probability distribution of small fluctuations around equilibrium can be written as
\[
P \propto
\exp\left[
-\frac{1}{2}
g_{ij}\,
dx^i dx^j
\right],
\]
where the Hessian of the relevant thermodynamic potential acts as the inverse covariance matrix of the fluctuations. In the entropy representation, the negative Hessian naturally provides a positive-definite metric whenever the entropy is locally concave. Thus, the metric measures the local stability of neighbouring equilibrium states under infinitesimal variations of the control parameters.\\

We stress upon the fact that our approach here must be viewed as an extension of the entropy representation to the manifold of intensive control parameters, rather than the conventional manifold of extensive state variables. Similar parameter-space constructions have been employed in several areas of thermodynamic geometry, including spin systems and black-hole thermodynamics, where quantities such as inverse temperature, chemical potentials, or coupling constants provide the most natural coordinates for describing equilibrium families of solutions.\\\\

Once the metric tensor \(g_{ij}\) is defined, the Riemann curvature tensor and the scalar curvature \(R\) can be computed using standard differential-geometric formulae. To determine the thermodynamic scalar curvature for a given thermodynamic metric, one follows the standard procedure of Riemannian geometry, with the metric defined on the space of equilibrium states. Let the metric be denoted by $g_{ij}(x^k)$, where $x^k$ represent the thermodynamic coordinates (such as entropy, charge, angular momentum, etc.). The first step is to compute the inverse metric $g^{ij}$. One then evaluates the Christoffel symbols accordingly as:
\[
\Gamma^{i}_{\;\;jk} = \frac{1}{2} g^{il} \left( \partial_j g_{lk} + \partial_k g_{lj} - \partial_l g_{jk} \right).
\]
Using these, the Riemann curvature tensor is constructed by the given formula:
\[
R^{i}_{\;\;jkl} = \partial_k \Gamma^{i}_{\;\;jl} - \partial_l \Gamma^{i}_{\;\;jk}
+ \Gamma^{i}_{\;\;km} \Gamma^{m}_{\;\;jl}
- \Gamma^{i}_{\;\;lm} \Gamma^{m}_{\;\;jk}.
\]
Contracting indices gives the Ricci tensor,
\[
R_{jl} = R^{i}_{\;\;jil},
\]
and finally the thermodynamic scalar curvature is obtained by a further contraction with the inverse metric,
\[
R = g^{jl} R_{jl}.
\]
The resulting scalar $R$ encodes information about thermodynamic interactions and phase transitions; in many systems, divergences of $R$ are associated with critical points, while its sign is often interpreted as reflecting the nature of the underlying microscopic interactions. The scalar curvature acts as a geometric diagnostic of interactions: \(R>0\) is often interpreted as corresponding to predominantly exclusion-like or repulsion-dominated correlations, \(R<0\) is commonly associated with clustering or attraction-like correlations, although this interpretation is heuristic and model dependent., and \(|R|\) is often interpreted to be proportional to the correlation volume \(\xi^d\), with \(\xi\) being the correlation length and \(d\) the dimensionality of the system \cite{ruppeiner2010interactions, janyszek1990ideal}. \\

In the context of the 1D Blume-Capel model, even though there is no genuine phase transition, the scalar curvature exhibits pseudo-critical peaks corresponding to maxima in the correlation length and also shows non zero scalar curvature, $R$ in certain cases for $q > 1$ regime. These peaks serve as geometric analogs of the crossovers observed in susceptibility and heat capacity. When the entropy is generalized via the Tsallis deformation, the geometry of the entropy surface is altered, leading to a modified curvature profile \(R(T)\). The shift of the curvature peak and its magnitude thus provide direct insight into how non-extensivity affects correlations, stability, and information content of the system in effect to change in the Tsallis parameter.

\subsection{Numerical Methodology}
To investigate the thermodynamic geometry of the one-dimensional Blume-Capel model by calculating the thermodynamic curvature scalar $R$, we must perform numerical calculations. Exact analytical computations of the scalar curvature become excruciatingly complex due to the highly non-linear nature of the generalized entropy and the requirement for high-order derivatives. To overcome this, we perform numerical calculations using \textbf{JAX}, a high-performance numerical computing python library that provides composable transformations, notably exact Automatic Differentiation (\textbf{AutoDiff}) and Just-In-Time (\textbf{JIT}) compilation. We do not use finite-difference methods since they are highly susceptible to truncation and floating-point rounding errors. When computing higher-order derivatives, \textbf{AutoDiff} evaluates derivatives analytically at the graph level. \\

While the standard differential-geometric procedure for calculating the scalar curvature $R$ involves computing the inverse metric, the Christoffel symbols $\Gamma^i_{jk}$, the Riemann curvature tensor, and subsequent tensor contractions, implementing this step-by-step tensor algebra numerically is computationally inefficient. For a two-dimensional thermodynamic manifold with the metric tensor components defined as $E = g_{\beta\beta}$, $F = g_{\beta J}$, and $G = g_{JJ}$, we can bypass the intermediate Christoffel symbols by utilizing the algebraically equivalent Brioschi formula. This approach is mathematically identical to the Christoffel symbol contraction but is vastly superior for vectorized automatic differentiation, as it compiles into a single, highly optimized computational graph. Calculating the Ricci scalar $R$ via the Brioschi formula requires the second derivatives of the metric components, which ultimately correspond to the fourth-order derivatives of the generalized entropy with respect to the thermodynamic parameters $(\beta, J)$. To ensure numerical stability during these extensive derivative calculations, we resort to enforcing all computational graphs in our code to operate in 64-bit precision as standard 32-bit floating-point precision is insufficient.  \\

We begin the calculation with the exact finite-size formulation of the model as has been derived in eq.\ref{eq:tsallis_numerical}. To ensure continuous differentiability during \textbf{AutoDiff}, we employ branch-less programming techniques ( smooth maximum and minimum functions) to consistently sort the remaining sub-dominant eigenvalues without breaking the computational graph. We implemented the entropy function $S(\beta, J)$ as a unified function capable of evaluating the Tsallis framework utilizing all three eigenvalues. \\

The computational pipeline for the geometry followed a three-step derivative process using JAX's transformation primitives. We extracted the metric components $E=g_{\beta\beta}$, $F=g_{\beta J}$, and $G=g_{JJ}$ by applying the Hessian operator directly to the entropy scalar function, returning the exact $2 \times 2$ covariance matrix at any point $(\beta, J)$. Then we applied the forward-mode Jacobian operator to the metric component vector x, after which, we applied a secondary Hessian operator to the metric components to yield the required higher-order cross-derivatives ($E_{JJ}, G_{\beta\beta}, F_{\beta J}$). Next, we substitute the computed metric components and their derivatives directly into the Brioschi determinant formula. The Brioschi formula directly expresses the Gaussian curvature $K$ entirely in terms of the metric components and their first and second partial derivatives:

\begin{equation}
K = \frac{1}{(EG-F^2)^2} \left( \begin{vmatrix} 
-\frac{1}{2}E_{JJ} + F_{\beta J} - \frac{1}{2}G_{\beta\beta} & \frac{1}{2}E_{\beta} & F_{\beta} - \frac{1}{2}E_J \\
F_J - \frac{1}{2}G_{\beta} & E & F \\
\frac{1}{2}G_J & F & G 
\end{vmatrix} 
- 
\begin{vmatrix} 
0 & \frac{1}{2}E_J & \frac{1}{2}G_{\beta} \\
\frac{1}{2}E_J & E & F \\
\frac{1}{2}G_{\beta} & F & G 
\end{vmatrix} \right)
\end{equation}

In this notation, the straight vertical bars denote the determinant of the $3 \times 3$ matrices, and the subscripts denote partial derivatives with respect to the thermodynamic coordinates (e.g., $E_{\beta} = \partial E / \partial \beta$ and $F_{\beta J} = \partial^2 F / \partial \beta \partial J$). The determinants of these matrices are efficiently computed using JAX's linear algebra modules. Finally, the thermodynamic Ricci scalar curvature $R$ for the two-dimensional parameter space is simply given by
$R = 2K$. \\

 For reproducibility, we summarize the numerical settings used in the curvature calculations. All computations were performed in dimensionless units with \(k_B=1\) and \(J=1\). The two anisotropy regimes considered were \(D=0.935\) for \(D<J\) and \(D=1.065\) for \(D>J\). 
The Tsallis indices used in the \(q>1\) sector were 
\(q=1.1,1.3,1.5,1.7\), while those used in the \(q<1\) sector were 
\(q=0.9,0.8,0.75,0.7\). 
To smoothly handle the Boltzmann-Gibbs limit ($n \to 1$ or $q \to 1$), we enforced a safe numerical convergence shift (evaluating at $1.00001$) to bypass zero-division errors while preserving the desired asymptotic accuracy. 
The temperature variable was defined as \(T=1/\beta\). We generated the high-resolution curvature profiles necessary to accurately capture the ultra-narrow pseudo-critical peaks, choosing $200,000$ grid points for every temperature interval of $1.0$, by vectorizing the entire Brioschi computational graph. Combined with JIT compilation, this allows the algorithm to compile the deep derivative graph into highly optimized machine code, yielding the exact scalar curvature across the entire parameter domain with high precision and computational efficiency. \\ \\

The values \(D=0.935\) and \(D=1.065\) were chosen to represent two physically distinct regimes of the model, namely the exchange-dominated regime (\(D<J\)) and the crystal-field-dominated regime (\(D>J\)). These values lie close to the crossover region \(D\approx J\), where the competition between magnetic and non-magnetic configurations is strongest and pseudo-critical signatures are most pronounced. The purpose of this choice is not to model a particular experimental material but rather to investigate how the thermodynamic geometry responds when the dominant microscopic configurations change. The system sizes \(N=60\) and \(N=600\) were selected to examine finite-size effects and to assess the robustness of the observed geometric features. The smaller system retains noticeable finite-size corrections, whereas the larger system serves as a near-bulk limit, allowing us to distinguish genuine geometric behaviour from finite-size artefacts.\\

Since the calculation is based on closed-form transfer-matrix expressions and automatic differentiation, no iterative nonlinear solver is involved. 
Accordingly, numerical convergence was assessed through grid-refinement tests rather than iterative stopping criteria. The results were checked by repeating the calculation on coarser and finer temperature grids and verifying that the pseudo-critical peak positions and curvature profiles remained stable. Numerical stability was further monitored by enabling 64-bit precision in JAX, using the dominant-eigenvalue factorized expression for the entropy, checking the determinant and eigenvalues of the metric tensor, and verifying representative curvature values against the direct Christoffel-symbol construction. These checks ensure that the reported curvature peaks are not artifacts of finite-difference errors, eigenvalue ordering, or floating-point instability.

\subsection{Physical and Informational Significance}

The thermodynamic metric also admits a direct interpretation within information geometry \cite{amari1985differential, crooks2007measuring}. The metric can be viewed as the Fisher information metric of the probability distribution \(p_i(\beta,J)\) associated with the system. For a family of equilibrium probability distributions $p_i(\beta,J)$, the Fisher information metric is given by
\[
g_{ij}^{(\mathrm{F})}
=
\sum_i
p_i
\frac{\partial \ln p_i}{\partial x^i}
\frac{\partial \ln p_i}{\partial x^j},
\]
which quantifies the statistical distinguishability between neighbouring equilibrium states. Under very general conditions, Hessian metrics generated from entropy or Massieu potentials are closely related to Fisher information geometry. In the present work, the scalar curvature derived from the entropy Hessian therefore retains a clear information-geometric interpretation: it characterizes the local structure of the statistical manifold and provides a measure of the effective correlation volume associated with the underlying equilibrium distribution.\\

We stress that our principal objective is to employ the scalar curvature as a geometric probe of the correlation structure and pseudo-critical behaviour induced by the generalized entropy deformation. The interpretation of the curvature therefore does not rely on the metric being the conventional Ruppeiner metric in the strict extensive-variable sense, but rather on its role as an entropy-Hessian/Fisher-information metric defined on the manifold of equilibrium control parameters. In this sense, the curvature \(R\) provides an information-theoretic measure of how sensitive the system is to changes in its thermodynamic parameters. For generalized statistics, the deformation parameter such as the \(q\) (Tsallis) modify the effective probability measure, thereby rescaling the Fisher information and reshaping the thermodynamic manifold. The resulting curvature captures the interplay between statistical weighting of microstates and macroscopic stability, establishing a deep connection between information theory, thermodynamics, and geometry.\\

Broadly speaking, the thermodynamic metric constructed from generalized entropy provides a unified geometric language to describe equilibrium fluctuations, correlation structure, and stability properties. When applied to the one-dimensional Blume-Capel model, it reveals how nonextensive deformations influence the geometry of the entropy surface and the corresponding thermodynamic curvature, offering a novel geometric perspective on generalized statistical mechanics.

\section{Thermodynamic Geometric analysis of the $1$-D Blume-Capel model}
\label{sec:thermodynamic}

\subsection{Tsallis entropy modifications}
Combining eq.\ref{eq:tsallis_numerical} with eq.\ref{eq:metric} we write the thermodynamic metric for the Tsallis modified Blume-Capel model in  \((\beta, J)\) co-ordinates given by:

\begin{equation}
g_{ij} = - \frac{\partial^2 S_{q}(\beta,J)}{\partial x^i \partial x^j}, \qquad x^i = (\beta, J),
\label{eq:Tsallis th. metric}
\end{equation}
where, $S_{q}(\beta,J)$ is given by eqtn.\ref{eq:tsallis_numerical}. Starting from the transfer matrix formulation of the model, the partition function was obtained and subsequently used to construct the Tsallis entropy, which depends explicitly on the inverse temperature $\beta$ and the interaction strength $J$, with the crystal–field anisotropy parameter $D$ controlling the relative population of the magnetic ($S=\pm1$) and non–magnetic ($S=0$) states. The thermodynamic metric was then defined as the negative Hessian of the generalized entropy with respect to the parameters $(\beta,J)$, thereby generating a two–dimensional thermodynamic manifold that captures the response of the entropy surface to fluctuations in temperature and interaction strength. From this metric the associated scalar curvature $R$ was calculated using the standard geometric construction of Riemannian thermodynamics. The curvature was evaluated as a function of temperature for different values of the Tsallis parameter $q$ and for two representative regimes of the anisotropy parameter, namely $D<J$ and $D>J$. Since the one–dimensional Blume–Capel model does not possess a true thermodynamic phase transition at finite temperature, the scalar curvature remains finite throughout the temperature range; however, pronounced peaks appear in $R(T)$ which signal regions of enhanced but finite correlations and therefore act as geometric indicators of pseudo–critical crossover behaviour. We then proceed to compute the thermodynamic scalar from the given metric which is a purely straightforward mathematical calculation. The thermodynamic scalar therefore obtained is plotted against inverse $\beta$ which is the characteristic temperature of the model denoted as $T$.\\

 \textbf{In order to analyze the influence of the crystal–field anisotropy on the thermodynamic geometry of the system, the results are presented separately for two representative parameter regimes characterized by the relative magnitude of the anisotropy parameter $D$ and the exchange interaction $J$. Specifically, we consider the cases $D<J$ ($D=0.935$, $J=1$) and $D>J$ ($D=1.065$, $J=1$). For each regime, the thermodynamic scalar curvature $R(T)$ is plotted as a function of temperature for different system sizes, which goes from $N=60$ to $N=1500$, in order to examine possible finite-size effects and the stability of the pseudo-critical features observed in the curvature profile. Since the scalar curvature is known to provide a geometric measure of the effective correlation volume of the system, its temperature dependence allows us to investigate how the underlying spin configurations and their statistical weights influence the correlation structure of the Blume–Capel chain.}\\

\textbf{The distinction between the regimes $D<J$ and $D>J$ has a clear physical interpretation in terms of the relative stability of the allowed spin states. When $D<J$, the exchange interaction dominates over the crystal-field anisotropy, and the energetically favorable configurations are the magnetic states $S=\pm1$, which tend to align due to the ferromagnetic coupling. In this regime the non-magnetic state $S=0$ is energetically disfavored and therefore occurs with relatively low probability, effectively behaving as a rare fluctuation in the spin configuration. Consequently, the dominant contributions to the thermodynamic properties arise from configurations composed primarily of $S=\pm1$ spins, while the $S=0$ states represent rare events that perturb the magnetic ordering. In contrast, when $D>J$, the crystal-field anisotropy becomes the dominant energy scale and favors the non-magnetic configuration $S=0$. In this situation the lattice tends to be populated predominantly by $S=0$ states, while the magnetic states $S=\pm1$ become energetically suppressed and appear only as thermally excited fluctuations. Thus, in this regime the $S=0$ configuration represents the most probable or “main” event, whereas the magnetic states act as rare excitations that introduce fluctuations into an otherwise non-magnetic background. By examining the curvature behaviour in these two distinct regimes, we can therefore understand how the thermodynamic geometry responds to changes in the dominant microscopic configurations and how the Tsallis non-extensive parameter modifies the statistical weighting of these common and rare events. With this physical interpretation in mind, we proceed to present the scalar curvature plots for the different parameter ranges.}\\

\textbf{In addition to the geometric analysis, we compare the thermodynamic scalar curvature with the specific heat capacity calculated within the same Tsallis framework in order to establish that the curvature extrema correspond to physically meaningful pseudo-critical crossovers identified by a conventional thermodynamic response function i.e. the scalar curvature. We compute the specific heat capacity within the same Tsallis framework of the 1D Blume capel model. And instead of evaluating the specific heat directly from the internal energy, we employ the thermodynamic identity
\[
C = T\left(\frac{\partial S_q}{\partial T}\right),
\]
where $S_q$ denotes the generalized Tsallis entropy constructed from the canonical partition function. This expression follows directly from the first law of thermodynamics, $dU = TdS$, for a system with fixed volume and particle number, implying
\[
C=\left(\frac{\partial U}{\partial T}\right)=T\left(\frac{\partial S}{\partial T}\right) = -\beta\left(\frac{\partial S}{\partial \beta}\right).
\]
Since the generalized entropy $S_q$ is itself obtained from the partition function, this procedure is mathematically equivalent to calculating the specific heat from the canonical partition function through the internal energy. Consequently, the specific heat and the thermodynamic scalar curvature are derived from the same underlying thermodynamic potential, providing a consistent framework for comparing conventional thermodynamic response functions with the geometric signatures of pseudo-critical behaviour.}

\section{Results and Discussion}
\label{sec:discussion}

In this section, we analyze the temperature dependence of the thermodynamic scalar curvature $R(T)$ for the one-dimensional Blume--Capel model within the Tsallis entropy formalism. The scalar curvature is computed from the thermodynamic metric defined as the negative Hessian of the generalized entropy with respect to the thermodynamic parameters $(\beta, J)$. Since the one-dimensional Blume--Capel model does not exhibit a true phase transition at finite temperature, the appearance of peaks in $R(T)$ is interpreted as a signature of pseudo-critical behaviour associated with enhanced but finite correlation lengths. Additionally, the peaks are also compared with the peaks obtained from the specific heat capacities of the spin model in both $D<J$ and $D>J$ regimes for both $q>1$ and $q<1$ each.\\

The analysis is divided into two regimes depending on the relative strength of the crystal field anisotropy $D$ and the exchange interaction $J$, namely $D<J$ and $D>J$. For each regime, we examine the behavior of the scalar curvature in the Boltzmann--Gibbs (BG) limit ($q=1$) and its deformation under the Tsallis entropy for $q>1$ and $q<1$. Finite-size effects are studied by comparing system sizes from $N=60$ to $N=1500$.

\subsection{$D<J$: Magnetically Dominated Regime}

When the exchange interaction dominates over the crystal field anisotropy ($D<J$), the magnetic states $S=\pm1$ constitute the most probable configurations, while the non-magnetic state $S=0$ corresponds to a rare event. 

 \begin{figure}[h!]
    \centering
    \includegraphics[width=0.99\textwidth]{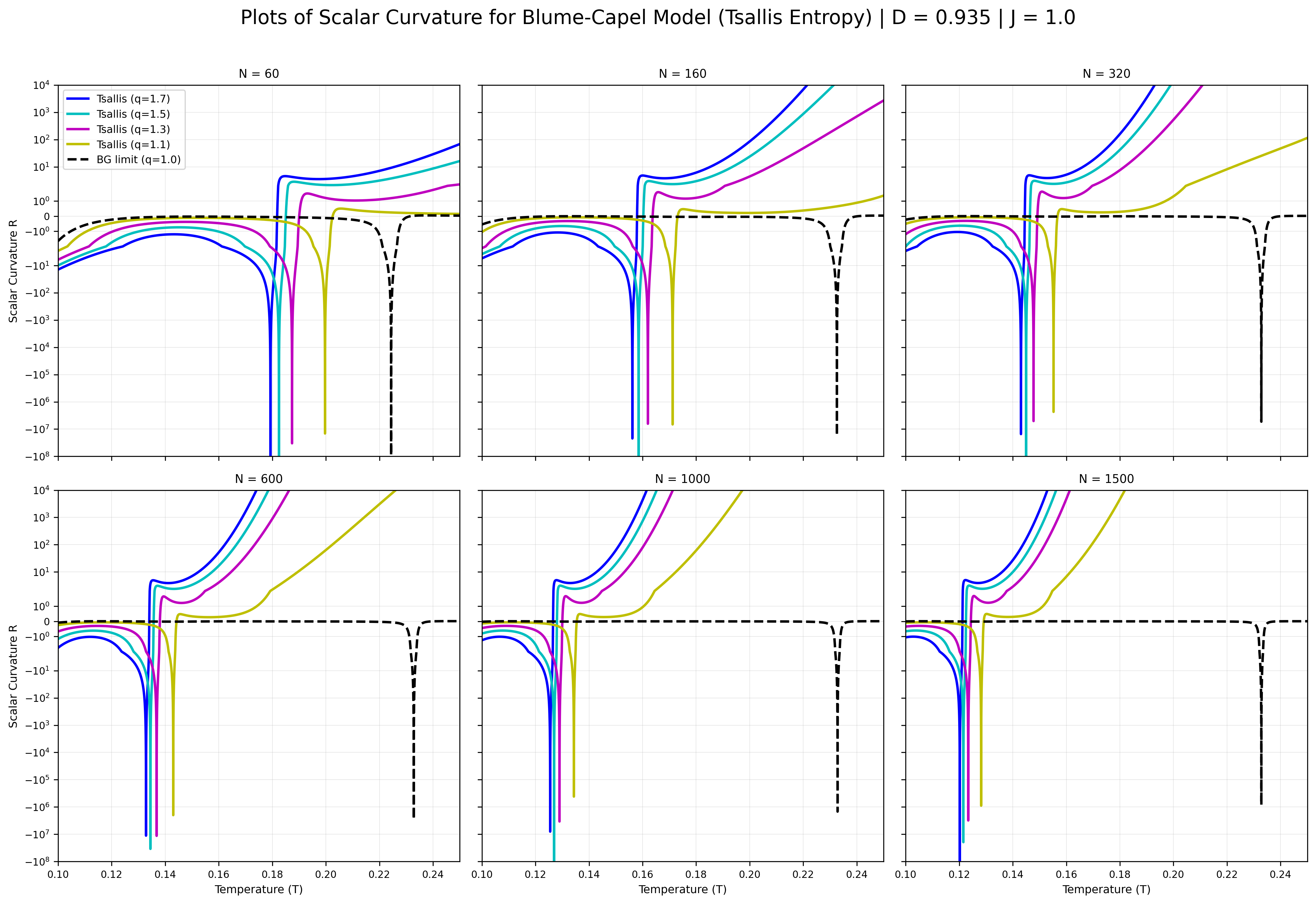}
    \caption{Temperature dependence of the thermodynamic scalar curvature \(R(T)\) for the one-dimensional Blume Capel model using the Tsallis entropy for \textbf{(D $<$ J)} $D=0.935, J=1$ and for $N=60,160,320,600,1000,1500$. For the \(q>1\) regime, the scalar curvature for the Boltzmann -Gibbs entropy (BG limit) ($q=1$) lies around $T=0.24$ and is negative, but for \(q>1\) the curvature peak though still negative shifts towards lower temperatures and the scalar curvature is non zero after the transition takes place for \(q>1\) case whereas it was zero in the previous BG limit.}
    \label{fig:01}
\end{figure}

\begin{figure}[h!]
    \centering
    \includegraphics[width=0.99\textwidth]{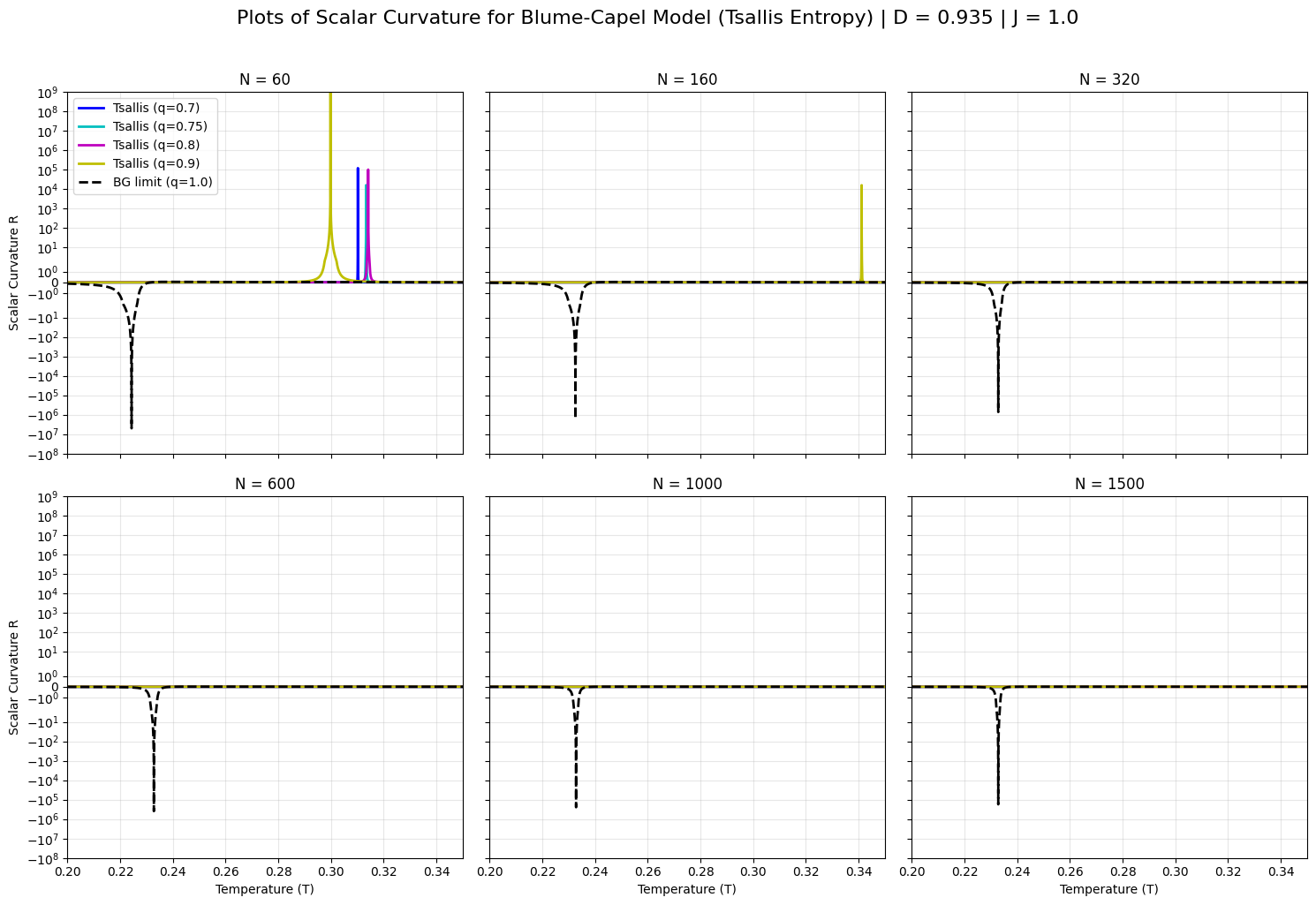}
    \caption{Temperature dependence of the thermodynamic scalar curvature \(R(T)\) for the one-dimensional Blume Capel model using the Tsallis entropy for \textbf{(D $<$ J)} $D=0.935, J=1$ and for $N=60,160,320,600,1000,1500$. For the \(q<1\) regime, the scalar curvature for the Boltzmann -Gibbs entropy (BG limit) also lies around $T=0.24$ and is negative but for \(q<1\) the curvature peak becomes positive and is seen to be suppressed for larger N values but they shift towards higher temperatures and here the scalar curvature is zero after the transition takes place in both the the BG limit and $q<1$ regime. }
    \label{fig:02}
\end{figure}

As we see from Fig.\ref{fig:01} and Fig.\ref{fig:02}\\

In the Boltzmann--Gibbs limit ($q=1$), the scalar curvature exhibits a pronounced negative peak around $T \approx 0.24$. The negative sign of $R$ indicates the dominance of attractive correlations associated with short-range ferromagnetic ordering. As the temperature increases beyond this crossover region, the curvature approaches zero, signaling the gradual loss of correlations and the emergence of a paramagnetic phase. This behavior is consistent with the absence of true criticality in one dimension.\\

\begin{figure}[h!]
    \centering
    \includegraphics[width=0.99\textwidth]{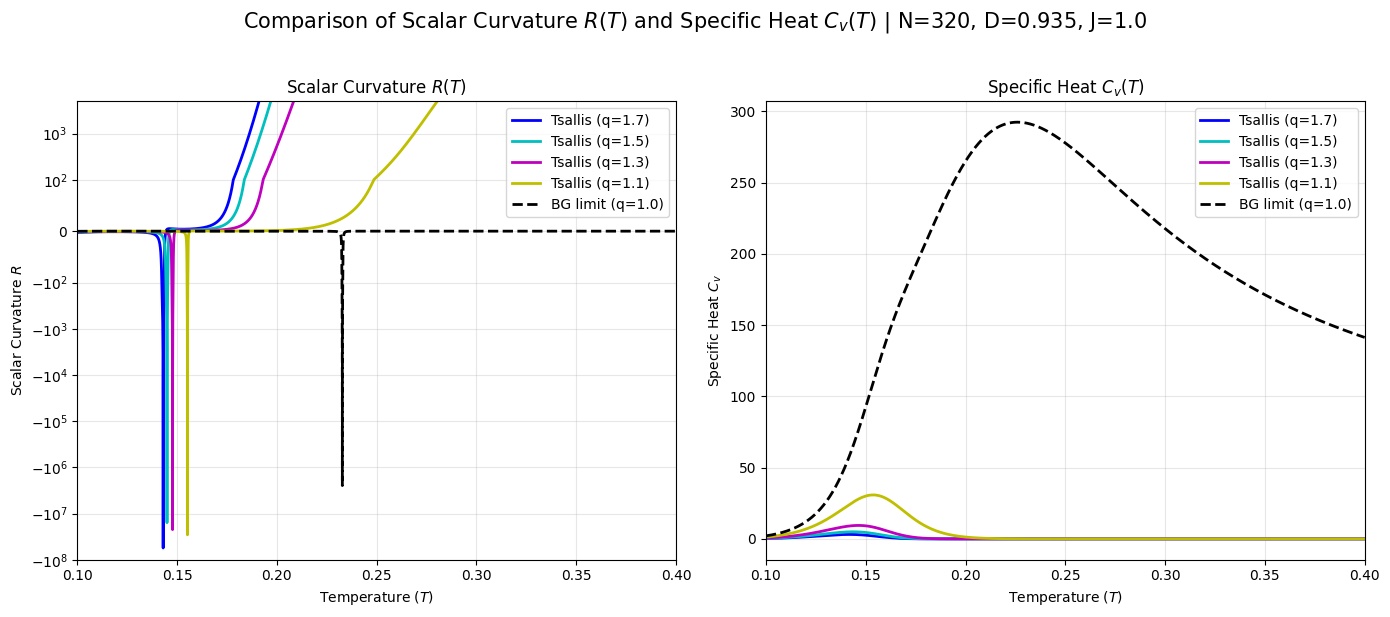}
    \caption{Temperature dependence of the scalar curvature \(R(T)\) and the specific heat capacity \(C_v (T)\) for the one-dimensional Blume Capel model using the Tsallis entropy for \textbf{(D $<$ J)} $D=0.935, J=1$ and for $N=320$ for the \textbf{q$>1$ case}. }
    \label{fig:03}
\end{figure}

\begin{figure}[h!]
    \centering
    \includegraphics[width=0.99\textwidth]{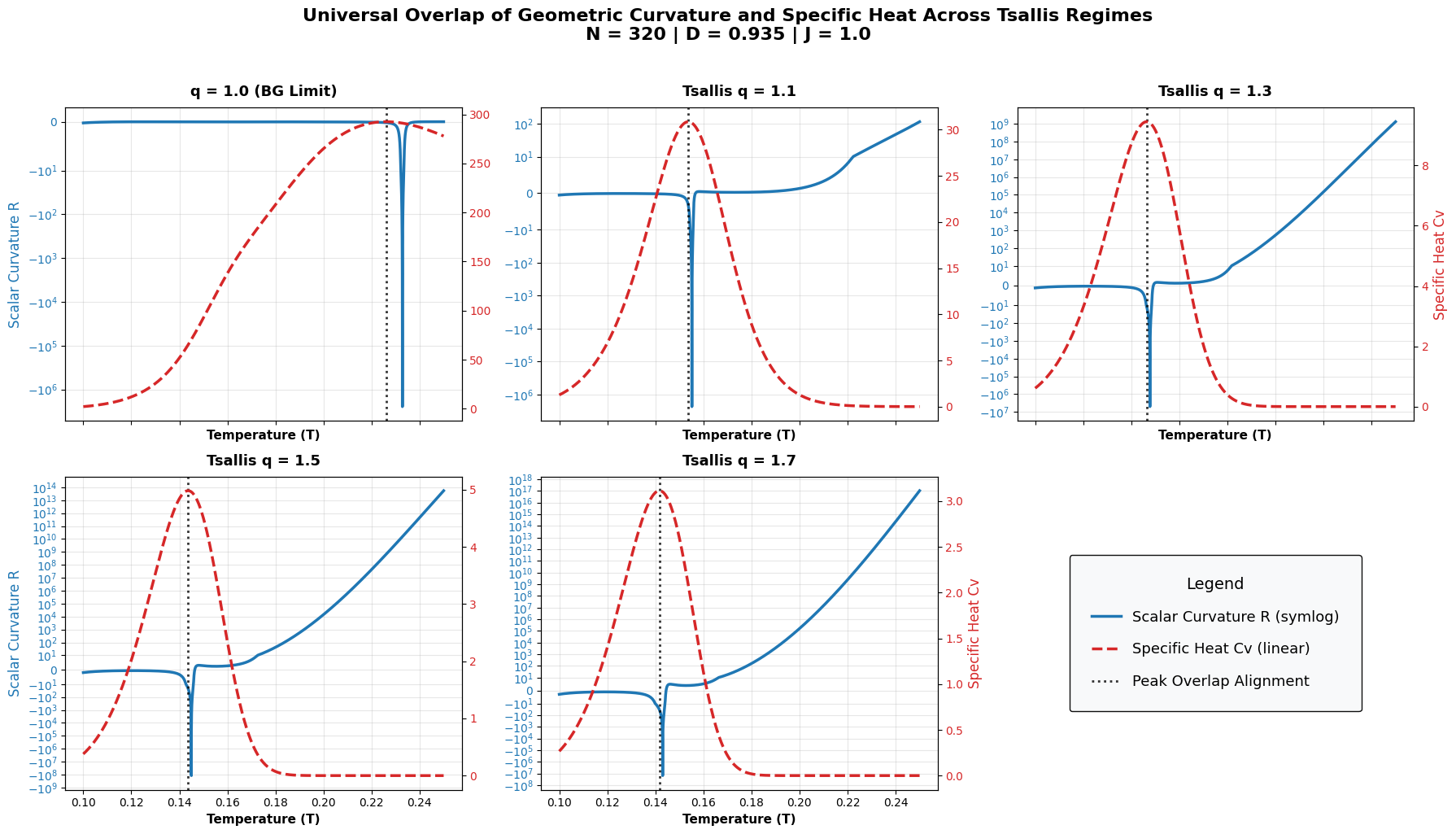}
    \caption{The overlap of scalar curvature peak with that of the peak of the sudden fluctuation in the specific heat capacity profile for \textbf{(D $<$ J)} $D=0.935, J=1$ and for $N=320$ for the \textbf{q$>1$ case}.  }
    \label{fig:04}
\end{figure}

For $q>1$ (Fig.\ref{fig:01}), where rare events are suppressed within the Tsallis framework, the contribution of the $S=0$ state is reduced. As a result, magnetic correlations are effectively reinforced. This is reflected in a shift of the curvature peak toward lower temperatures while maintaining its negative sign. Moreover, unlike the Boltzmann--Gibbs case, the scalar curvature remains nonzero even beyond the pseudo-transition temperature, indicating the persistence of residual correlations induced by the nonextensive deformation. The behaviour of the curvature peaks is also avidly reflected when these curvature profiles are compared with that of the specific heat capacities of the spin model for the $q>1$ case as can be seen from the figures Fig.\ref{fig:03} and Fig.\ref{fig:04} where we clearly see that for approximately the same temperature at which the scalar curvature shows peaks, there is corresponding fluctuation in the specific heat capacity profile which quickly flattens out for the rest of the temperature profile. The slight discrepancy in the positions of the curvature peaks and that of the fluctuations in the specific heat profile  is properly depicted in Table \ref{tab:Dless_qgt_overlap}. This discrepancy is a common occurrence in fisher-information like metrics and is often seen for other instances particularly in applications of metric such as the Ruppeiner metric. Difficulties as such were adequately dealt with, when, the legendre invariant geometrothermodynamic metric was proposed. However, for the purpose of our work here the approximate overlap of the specific heat fluctuation with that of the curvature profile is good enough to embolden the validity of our results.\\

\begin{table}[h!]
\centering
\small
\setlength{\tabcolsep}{4pt}

\begin{minipage}{0.48\textwidth}
\centering
\textbf{\(q>1\)}
\vspace{0.2cm}

\begin{tabular}{c|cccc}
\hline
\(N \backslash q\) & \(1.10\) & \(1.30\) & \(1.50\) & \(1.70\) \\
\hline
60   & 0.19967 & 0.18737 & 0.18247 & 0.17929 \\
160  & 0.17122 & 0.16198 & 0.15852 & 0.15619 \\
320  & 0.15521 & 0.14775 & 0.14497 & 0.14304 \\
600  & 0.14299 & 0.13679 & 0.13447 & 0.13282 \\
1000 & 0.13435 & 0.12897 & 0.12695 & 0.12549 \\
1500 & 0.12817 & 0.12336 & 0.12153 & 0.12020 \\
\hline
\end{tabular}
\end{minipage}
\hfill
\begin{minipage}{0.48\textwidth}
\centering
\textbf{\(q<1\)}
\vspace{0.2cm}

\begin{tabular}{c|cccc}
\hline
\(N \backslash q\) & \(0.70\) & \(0.75\) & \(0.80\) & \(0.90\) \\
\hline
60   & 0.31021 & 0.31333 & 0.31411 & 0.29981 \\
160  & --      & --      & --      & 0.34116 \\
320  & --      & --      & --      & --      \\
600  & --      & --      & --      & --      \\
1000 & --      & --      & --      & --      \\
1500 & --      & --      & --      & --      \\
\hline
\end{tabular}
\end{minipage}

\caption{Peak positions \(T_R\) of the scalar curvature \(R(T)\) for \(D=0.935\), \(J=1\), in the \(q>1\) and \(q<1\) Tsallis regimes. Here ``--'' indicates that no well-defined scalar-curvature peak was detected within the scanned temperature range.}
\label{tab:Dless_qgt_qlt_Rpeaks}
\end{table}

\clearpage

\begin{figure}[h!]
    \centering
    \includegraphics[width=0.99\textwidth]{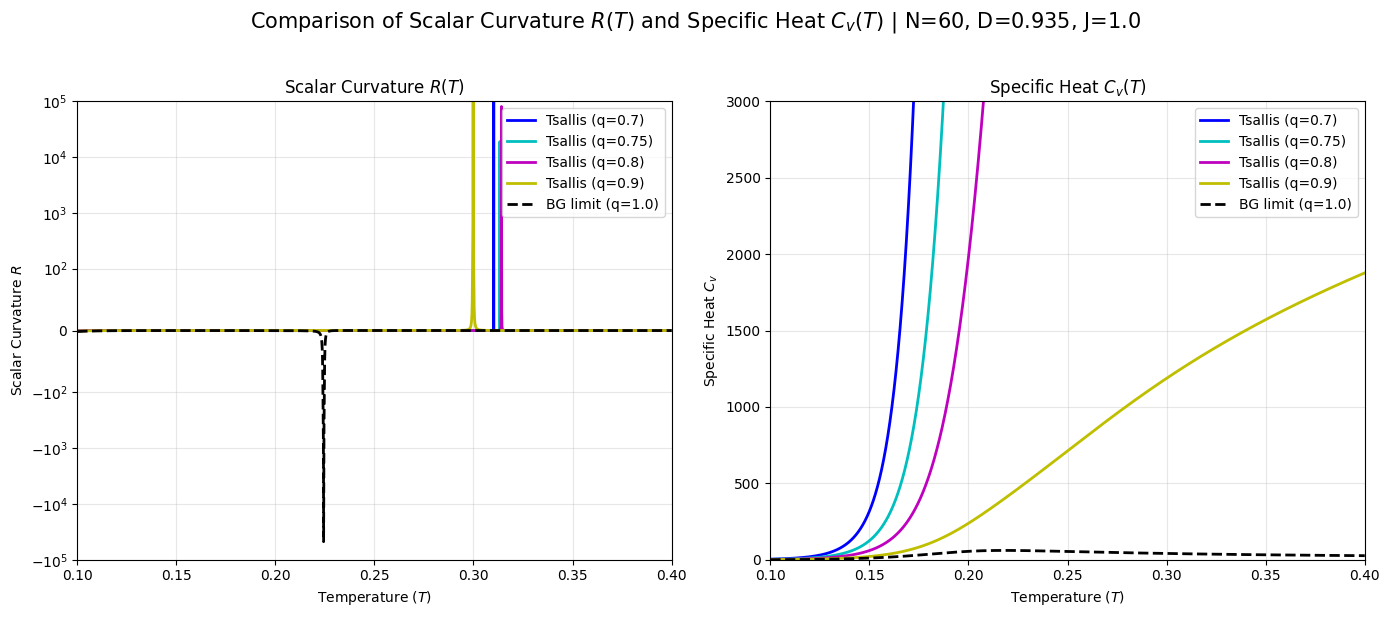}
    \caption{Temperature dependence of the scalar curvature \(R(T)\) and the specific heat capacity \(C_v (T)\) for the one-dimensional Blume Capel model using the Tsallis entropy for \textbf{(D $<$ J)} $D=0.935, J=1$ and for $N=60$ for the \textbf{q$<1$ case}. }
    \label{fig:05}
\end{figure}

In contrast, for $q<1$ (Fig.\ref{fig:02}), rare events are enhanced, leading to an increased population of the $S=0$ state. This enhancement disrupts magnetic clustering and weakens ferromagnetic correlations. Consequently, the curvature peak becomes positive and shifts toward higher temperatures, indicating the emergence of repulsion-like or exclusion-dominated correlations. Furthermore, the scalar curvature rapidly approaches zero beyond the crossover region, suggesting a more efficient suppression of correlations when rare states are amplified. The peaks quickly diminish with the increasing size of $N$ as can be seen from the Fig.\ref{fig:05} and we also see that unlike the $q>1$ case here the curvature profiles do not correspond to any fluctuation in the specific heat capacity profiles as can be seen from Fig. and this shows that the system is highly unstable in this regime and cannot be appropriately explained in the current information geometric measures.

\begin{table}[h!]
\centering

\begin{tabular}{c|ccc}
\hline
\(q\) & \(T_R\) & \(T_{C_v}\) & \(\Delta T=|T_R-T_{C_v}|\) \\
\hline
1.00 & 0.23282 & 0.22601 & 0.006815 \\
1.10     & 0.15522 & 0.15371 & 0.001510 \\
1.30     & 0.14775 & 0.14638 & 0.001370 \\
1.50     & 0.14497 & 0.14362 & 0.001345 \\
1.70     & 0.14305 & 0.14172 & 0.001330 \\
\hline
\end{tabular}
\caption{Comparison of scalar-curvature peak position \(T_R\) and specific-heat maximum \(T_{C_v}\) for \(D=0.935\), \(J=1\), and \(N=320\).}
\label{tab:Dless_qgt_overlap}
\end{table}

\subsection{$D>J$: Crystal-Field-Dominated Regime}

\begin{figure}[h!]
    \centering
    \includegraphics[width=0.99\textwidth]{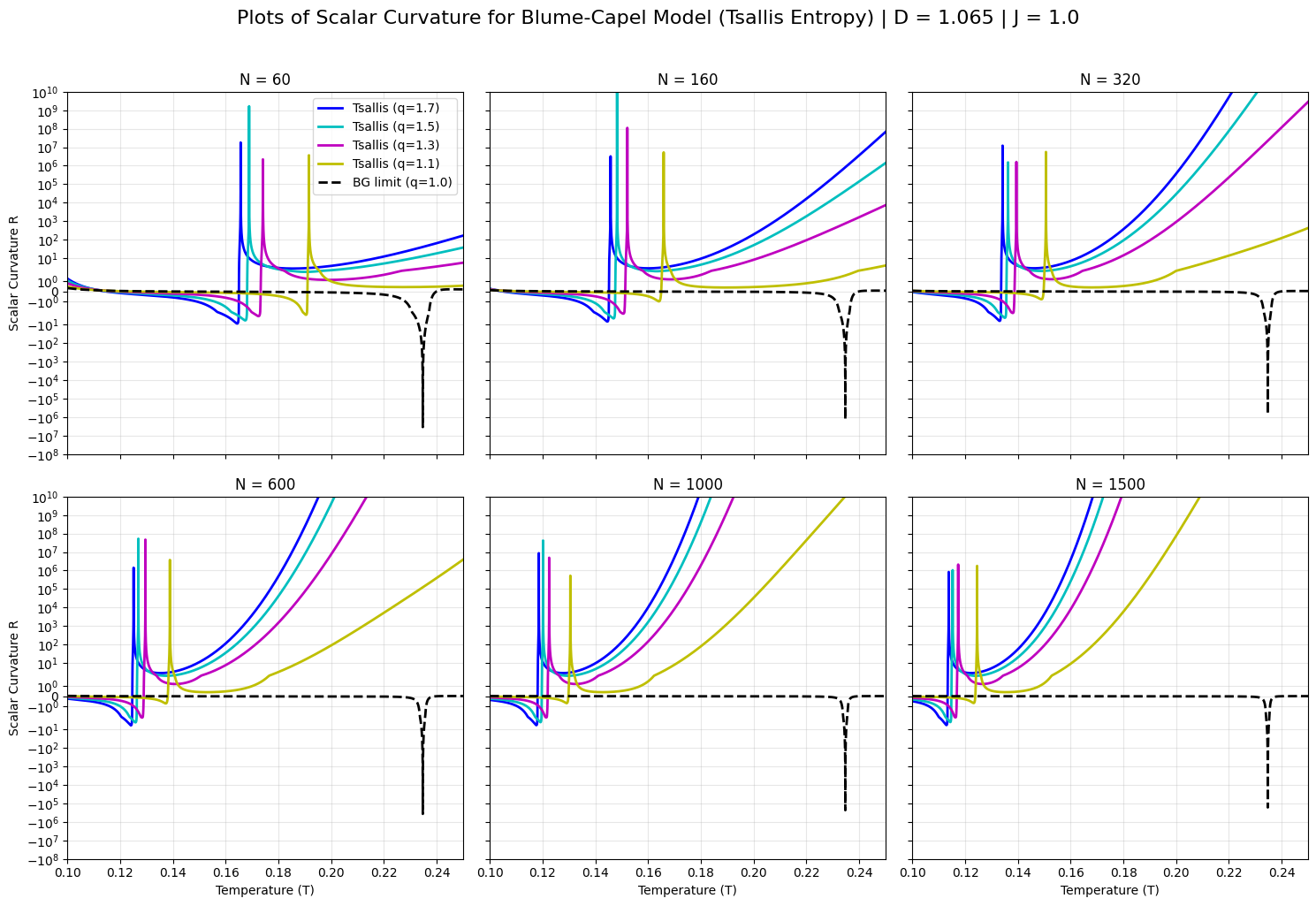}
    \caption{ Temperature dependence of the thermodynamic scalar curvature \(R(T)\) for the one-dimensional Blume Capel model using the Tsallis entropy for \textbf{(D $>$ J)} $D=1.065, J=1$ and for $N=60,160,320,600,1000,1500$. For the \(q>1\) regime, the scalar curvature for the Boltzmann -Gibbs entropy (BG limit) ($q=1$) lies around $T=0.24$ and is negative, but for \(q>1\) the curvature peak becomes positive and shifts towards lower temperatures and the scalar curvature is non zero after the transition takes place for \(q>1\) case whereas it was zero in the previous BG limit.}
    \label{fig:06}
\end{figure}

\begin{figure}[h!]
    \centering
    \includegraphics[width=0.99\textwidth]{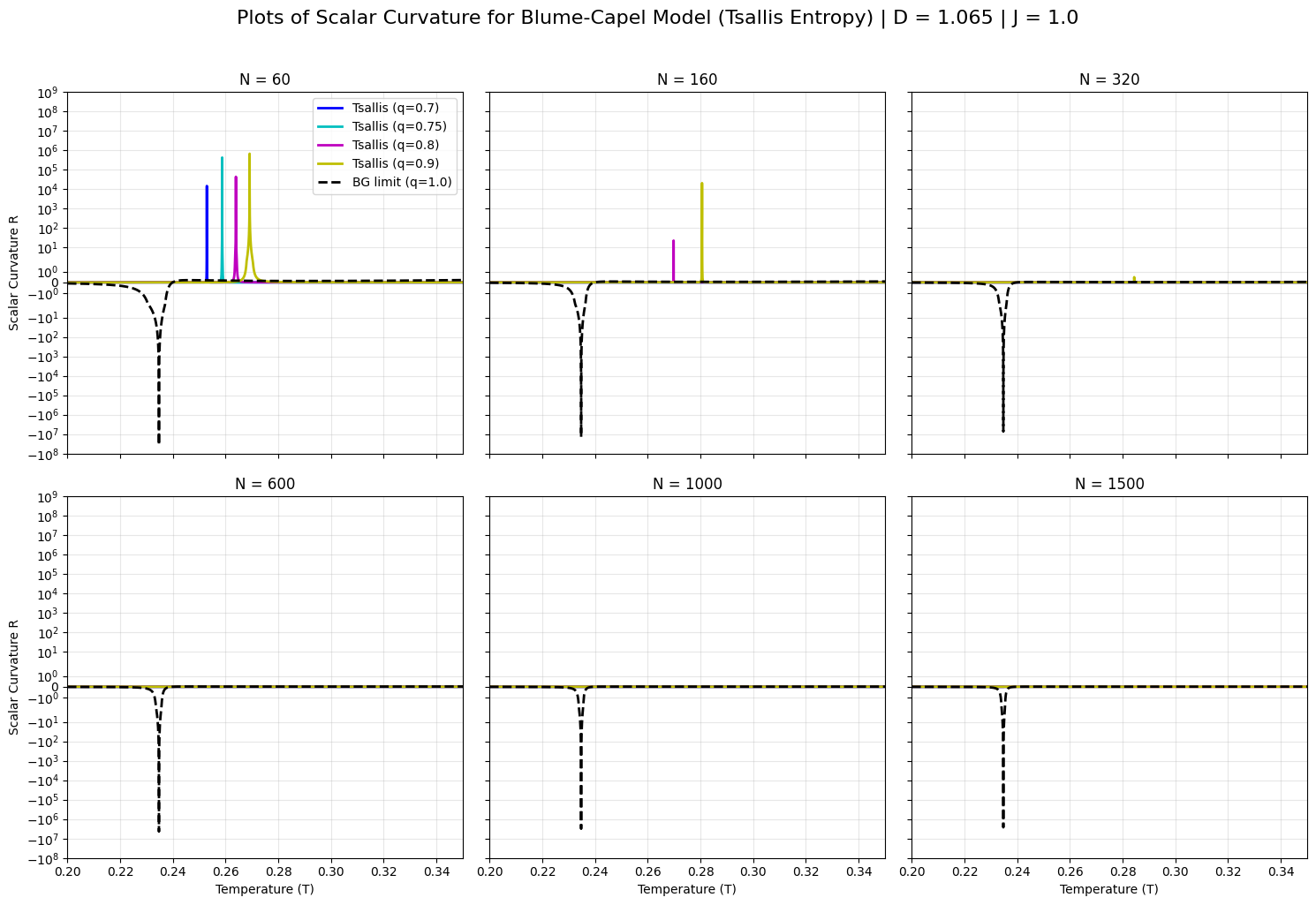}
    \caption{Temperature dependence of the thermodynamic scalar curvature \(R(T)\) for the one-dimensional Blume Capel model using the Tsallis entropy for \textbf{(D $>$ J)} $D=1.065, J=1$ and for $N=60,160,320,600,1000,1500$. For the \(q<1\) regime, the scalar curvature for the Boltzmann -Gibbs entropy (BG limit) also lies around $T=0.24$ and is negative but for \(q<1\) the curvature peak becomes positive and is seen to be suppressed for larger N values but they shift towards higher temperatures and here the scalar curvature is zero after the transition takes place in both the the BG limit and $q<1$ regime. }
    \label{fig:07}
\end{figure}

In the regime $D>J$, the crystal field anisotropy favors the non-magnetic $S=0$ state, which becomes the dominant configuration, while the magnetic states $S=\pm1$ are rare. As we see from Fig.\ref{fig:06} and Fig.\ref{fig:07}: \\

\begin{figure}[h!]
    \centering
    \includegraphics[width=0.99\textwidth]{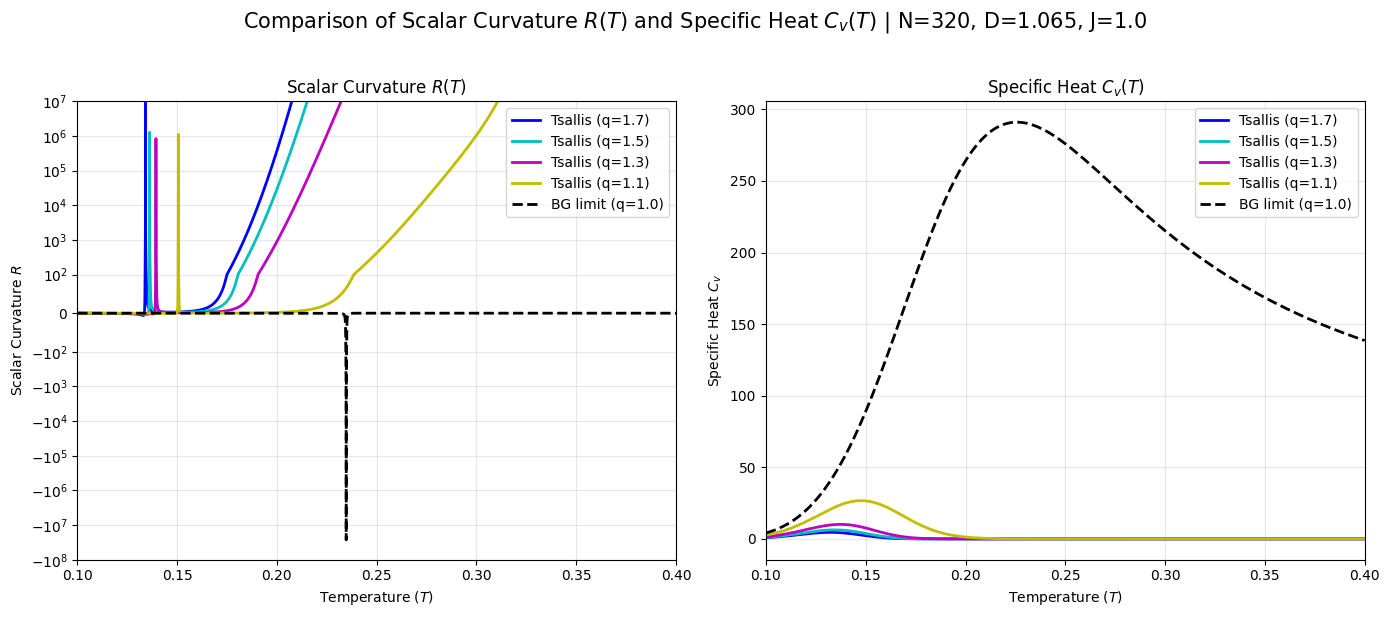}
    \caption{Temperature dependence of the scalar curvature \(R(T)\) and the specific heat capacity \(C_v (T)\) for the one-dimensional Blume Capel model using the Tsallis entropy for \textbf{(D $>$ J)} $D=1.065, J=1$ and for $N=320$ for the \textbf{q$>1$ case}. }
    \label{fig:08}
\end{figure}

\begin{figure}[h!]
    \centering
    \includegraphics[width=0.99\textwidth]{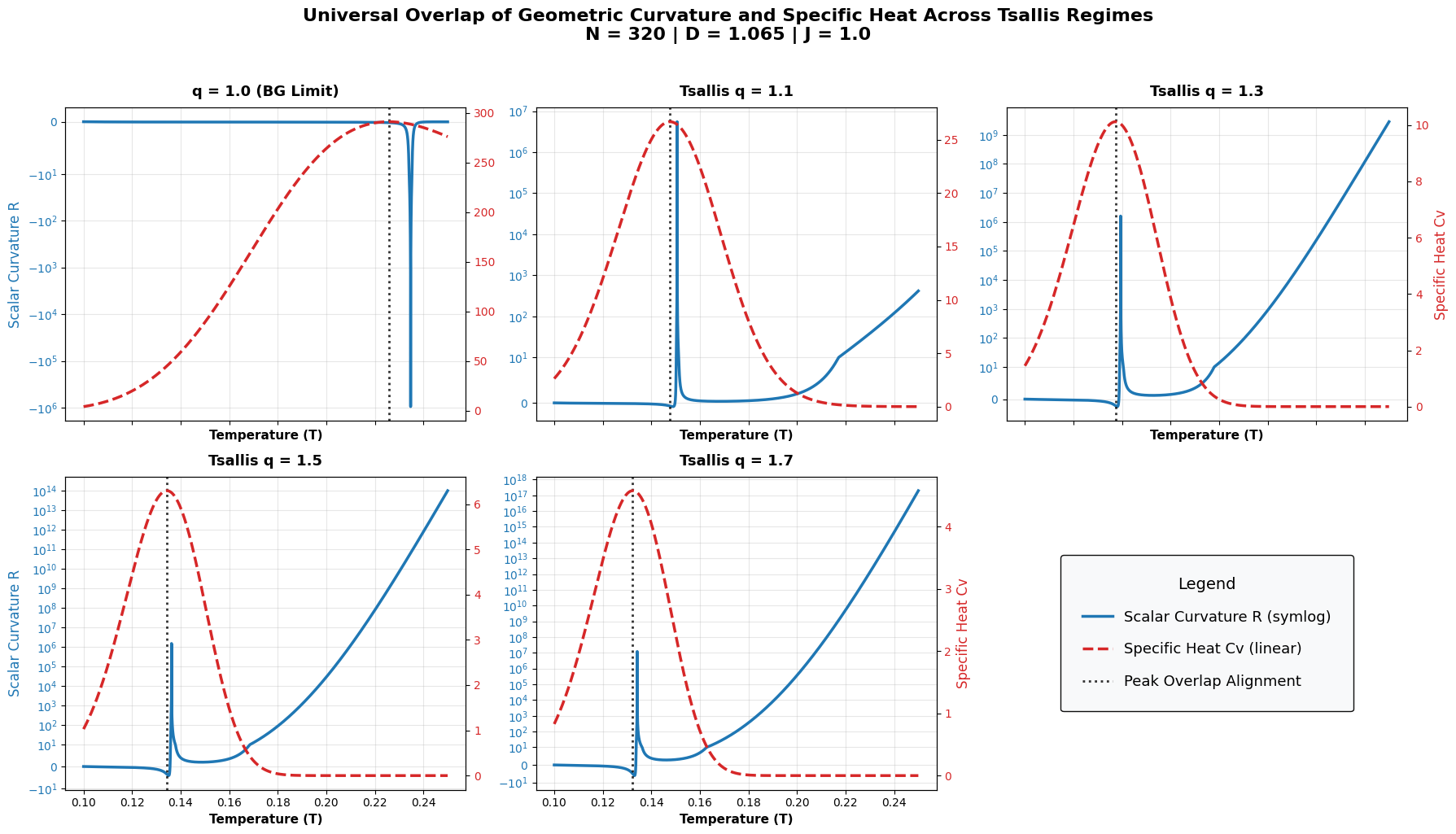}
    \caption{The overlap of scalar curvature peak with that of the peak of the sudden fluctuation in the specific heat capacity profile for \textbf{(D $>$ J)} $D=1.065, J=1$ and for $N=320$ for the \textbf{q$>1$ case}.  }
    \label{fig:09}
\end{figure}

In the Boltzmann--Gibbs limit, the scalar curvature again displays a negative peak near $T \approx 0.24$, indicating residual magnetic correlations arising from thermal excitation of the $S=\pm1$ states. As the temperature increases, the curvature approaches zero, consistent with a paramagnetic regime dominated by uncorrelated fluctuations.\\
\clearpage 
For $q>1$, the suppression of rare magnetic states further stabilizes the non-magnetic phase. In this case, the curvature peak changes sign and becomes positive, while shifting toward lower temperatures. The positive curvature indicates that the dominant correlations are no longer clustering-type but instead resemble exclusion-driven or repulsive interactions associated with the prevalence of the $S=0$ state. As observed in the magnetically dominated regime, the scalar curvature remains nonzero beyond the pseudo-transition, highlighting the persistence of correlations generated by the Tsallis deformation. As was evident in the previous case for $D<J$, the shifting behaviour of the positive curvature peaks is also avidly reflected when these curvature profiles are compared with that of the specific heat capacities of the spin model for the $q>1$ case as can be seen from the figures Fig.\ref{fig:08} and Fig.\ref{fig:09} where we clearly see that for approximately the same temperature at which the scalar curvature shows peaks, there is corresponding fluctuation in the specific heat capacity profile which quickly flattens out for the rest of the temperature profile. The slight discrepancy in the curvature peak and the specific heat fluctuation is avidely depicted in Table \ref{tab:Dgreat_qgt_overlap}.\\

\begin{table}[htbp]
\centering
\small
\setlength{\tabcolsep}{4pt}

\begin{minipage}{0.48\textwidth}
\centering
\textbf{\(q>1\)}
\vspace{0.2cm}

\begin{tabular}{c|cccc}
\hline
\(N \backslash q\) & \(1.10\) & \(1.30\) & \(1.50\) & \(1.70\) \\
\hline
60   & 0.19158 & 0.17413 & 0.16889 & 0.16569 \\
160  & 0.16588 & 0.15211 & 0.14822 & 0.14576 \\
320  & 0.15064 & 0.13943 & 0.13626 & 0.13419 \\
600  & 0.13886 & 0.12956 & 0.12690 & 0.12511 \\
1000 & 0.13050 & 0.12247 & 0.12015 & 0.11855 \\
1500 & 0.12453 & 0.11736 & 0.11526 & 0.11379 \\
\hline
\end{tabular}
\end{minipage}
\hfill
\begin{minipage}{0.48\textwidth}
\centering
\textbf{\(q<1\)}
\vspace{0.2cm}

\begin{tabular}{c|cccc}
\hline
\(N \backslash q\) & \(0.70\) & \(0.75\) & \(0.80\) & \(0.90\) \\
\hline
60   & 0.25292 & 0.25873 & 0.26394 & 0.26914 \\
160  & --      & --      & 0.26982 & 0.28061 \\
320  & --      & --      & --      & --      \\
600  & --      & --      & --      & --      \\
1000 & --      & --      & --      & --      \\
1500 & --      & --      & --      & --      \\
\hline
\end{tabular}
\end{minipage}

\caption{Peak positions \(T_R\) of the scalar curvature \(R(T)\) for \(D=1.065\), \(J=1\), in the \(q>1\) and \(q<1\) Tsallis regimes. Here ``--'' indicates that no well-defined scalar-curvature peak was detected within the scanned temperature range.}
\label{tab:Dgreat_qgt_qlt_Rpeaks}
\end{table}

\begin{figure}[h!]
    \centering
    \includegraphics[width=0.99\textwidth]{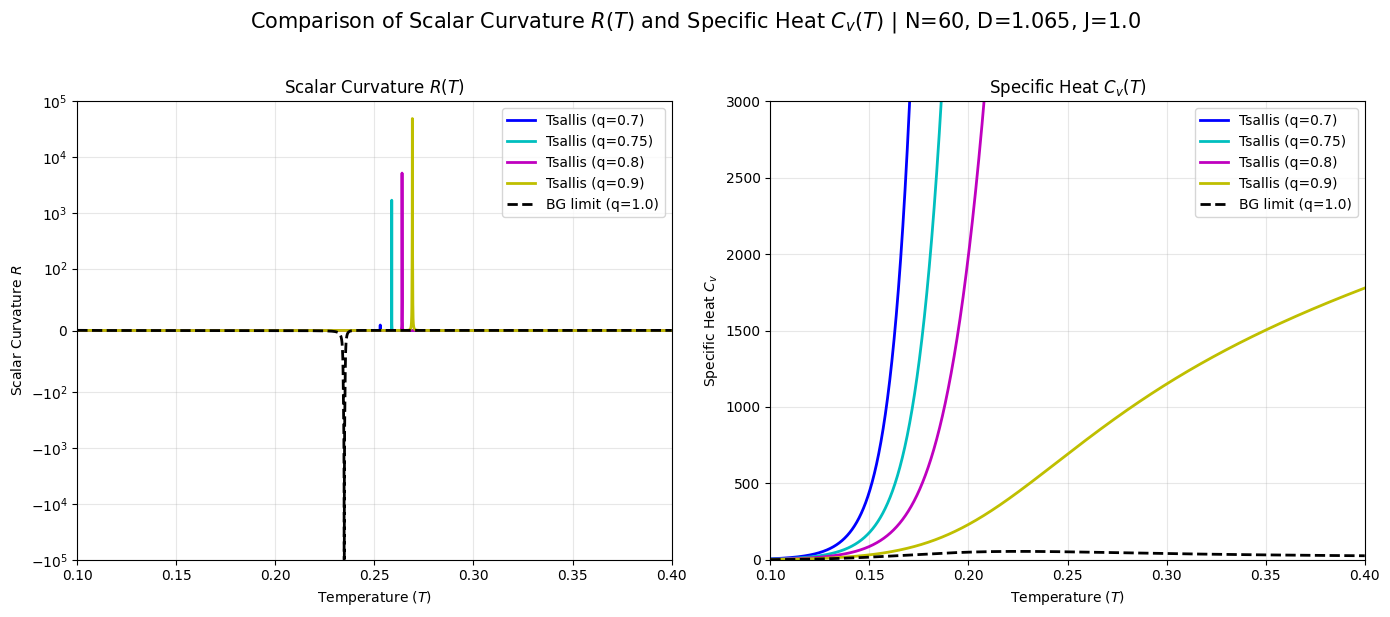}
    \caption{Temperature dependence of the scalar curvature \(R(T)\) and the specific heat capacity \(C_v (T)\) for the one-dimensional Blume Capel model using the Tsallis entropy for \textbf{(D $>$ J)} $D=1.065, J=1$ and for $N=60$ for the \textbf{q$<1$ case}. }
    \label{fig:10}
\end{figure}

For $q<1$, the enhancement of rare magnetic states partially restores magnetic fluctuations. However, these fluctuations do not develop into strong collective behaviour. As a result, the curvature peak is significantly suppressed or completely absent, and the scalar curvature remains close to zero over the entire temperature range. This indicates the absence of a dominant correlation structure when rare magnetic states are excessively weighted. As was observed in the $D<J$ case previously, the peaks here also quickly diminish with the increasing size of $N$ as can be seen from the Fig.\ref{fig:10} and we also see that unlike the $q>1$ case here the curvature profiles do not correspond to any fluctuation in the specific heat capacity profiles as can be seen from Fig. and this shows that the system is highly unstable in this regime and cannot be appropriately explained in the current information geometric measures.

\begin{table}[htbp]
\centering
\caption{Comparison of scalar-curvature peak position \(T_R\) and specific-heat maximum \(T_{C_v}\) for \(D=1.065\), \(J=1\), and \(N=320\).}
\label{tab:Dgreat_qgt_overlap}
\begin{tabular}{c|ccc}
\hline
\(q\) & \(T_R\) & \(T_{C_v}\) & \(\Delta T=|T_R-T_{C_v}|\) \\
\hline
1.000001 & 0.23474 & 0.22571 & 0.009035 \\
1.10     & 0.15064 & 0.14768 & 0.002960 \\
1.30     & 0.13943 & 0.13731 & 0.002120 \\
1.50     & 0.13626 & 0.13433 & 0.001925 \\
1.70     & 0.13419 & 0.13236 & 0.001835 \\
\hline
\end{tabular}
\end{table}

\clearpage

The observed dependence of the thermodynamic scalar curvature on the nonextensive parameter $q$ can be understood from the statistical weighting mechanism inherent in the Tsallis entropy,
\begin{equation}
S_q=\frac{1-\sum_i p_i^q}{q-1}.
\end{equation}
For configurations with small probabilities ($p_i\ll1$), one has $p_i^q<p_i$ when $q>1$, whereas $p_i^q>p_i$ when $q<1$. Consequently, values of $q>1$ reduce the entropic contribution of rare configurations, while values of $q<1$ enhance their contribution. Since the thermodynamic metric is constructed from the Hessian of the generalized entropy,
\begin{equation}
g_{ij}=-\frac{\partial^2 S_q}{\partial x^i \partial x^j},
\end{equation}
with $x^i=(\beta,J)$, any modification of the statistical weights directly alters the local curvature properties of the entropy surface and therefore affects the resulting thermodynamic scalar curvature.\\

The microscopic origin of this effect depends on the competition between the exchange interaction $J$ and the crystal-field anisotropy $D$. In the regime $D<J$, the ferromagnetic interaction favors the $S=\pm1$ states, while the $S=0$ state occurs with comparatively lower probability and therefore acts as a rare configuration. In contrast, for $D>J$, the crystal field favors the $S=0$ state and the magnetic states $S=\pm1$ become the less probable configurations. The Tsallis parameter $q$ therefore modifies the relative importance of these subdominant states in the entropy calculation. Enhancement of their contribution for $q<1$ increases the competition among microscopic configurations, producing stronger geometric responses and more pronounced curvature extrema. Conversely, for $q>1$, the suppression of these configurations reduces their influence on the entropy landscape, leading to smoother geometric behavior and modified curvature profiles.\\

From this perspective, the changes observed in the scalar-curvature profiles are not merely numerical consequences of varying $q$, but reflect the manner in which nonextensive statistics reshapes the underlying entropy landscape through the reweighting of low-probability configurations. The curvature therefore provides a geometric measure of the sensitivity of the generalized entropy to these modified statistical fluctuations.

\subsection{Finite-Size Effects and Pseudo-Criticality}

A comparison between system sizes starting from $N=60$ to $N=1500$ shows that increasing $N$ has drastic effects particularly for the $q<1$ regime of the Tsallis modified scalar curvature where the correlation peaks are completely suppressed for large enough $N$ and which shows that increasing system size reduces the effects of the rare events considerably whereas for the  $q>1$ regime the correlations survive even after the pseudo transition temperature has been reached.  The Tsallis deformation modifies the strength and persistence of correlations in accordance with the  size of the system for the spin model. We would like to stress here in particular that, within our analysis, the thermodynamic scalar curvature should primarily be regarded as a geometric measure of the structure and strength of thermodynamic correlations encoded in the entropy manifold. The sign of the curvature is therefore interpreted only in the conventional heuristic sense commonly employed in the literature and not as a universal criterion for identifying the microscopic nature of interactions.\\

The finite-size analysis in the magnetically dominated regime \(D<J\) shows a clear distinction between the \(q>1\) and \(q<1\) sectors. For \(q>1\), the scalar-curvature anomaly remains visible over the full range of system sizes \(N=60,160,320,600,1000,1500\). The extracted peak positions \(T_R\) show a systematic shift toward lower temperature as \(N\) increases. For example, for \(q=1.10\), the curvature peak moves from \(T_R=0.19967\) at \(N=60\) to \(T_R=0.12817\) at \(N=1500\). Similar monotonic shifts are observed for the other \(q>1\) values. This indicates that the \(q>1\) Tsallis deformation produces a robust pseudo-critical geometric signature whose location evolves systematically with system size. In contrast, for \(q<1\), well-defined curvature peaks are found only for small system sizes. For larger \(N\), the peaks are strongly suppressed or no longer detectable within the scanned temperature interval. This suggests that the \(q<1\) enhancement of rare \(S=0\) configurations weakens the collective geometric response in the large-\(N\) direction. \\ \\
A similar but physically distinct finite-size trend is observed in the crystal-field-dominated regime \(D>J\). For \(q>1\), the scalar-curvature anomaly persists for all system sizes considered, again shifting toward lower temperature as \(N\) increases. For instance, at \(q=1.10\), the peak position changes from \(T_R=0.19158\) for \(N=60\) to \(T_R=0.12453\) for \(N=1500\). This trend is consistent across the other \(q>1\) values and demonstrates that the curvature anomaly remains a stable pseudo-critical indicator even when the non-magnetic \(S=0\) state is the dominant microscopic configuration. In the \(q<1\) sector, however, the curvature peaks are again rapidly suppressed with increasing \(N\), and for sufficiently large systems no well-defined peak is detected. Thus, while the sign and magnitude of the curvature response depend on whether the dominant configurations are magnetic or non-magnetic, the finite-size behaviour shows a common pattern: the \(q>1\) sector supports robust curvature anomalies, whereas the \(q<1\) sector produces strongly finite-size-sensitive features.\\
 
The finite-size results therefore show that the curvature anomalies observed in this work are pseudo-critical crossover signatures rather than true thermodynamic-limit divergences. This is consistent with the absence of a finite-temperature phase transition in the one-dimensional Blume--Capel model. The range \(N=60\) to \(N=1500\) is sufficient to reveal the dominant finite-size trends: persistent and systematically shifting curvature anomalies for \(q>1\), and rapid suppression of the curvature response for \(q<1\). The close agreement between the scalar-curvature peak position \(T_R\) and the specific-heat maximum \(T_{C_v}\) for \(N=320\) further supports the interpretation that the robust \(q>1\) curvature anomaly identifies the same pseudo-critical crossover region as the conventional thermodynamic response.

Overall, the results demonstrate that Tsallis entropy reshapes the geometric structure of the thermodynamic state space by selectively reweighting rare and common microstates. The sign, magnitude, and temperature location of the scalar curvature peak are governed by the interplay between the dominant spin configurations and the nonextensive parameter $q$. While the Boltzmann--Gibbs limit reflects conventional short-range correlations, deviations from extensivity induce effective long-range correlations that persist beyond the pseudo-transition region. Nevertheless, the absence of curvature divergence reaffirms the pseudo-critical nature of these geometric signatures in the one-dimensional Blume--Capel model.

\section{Conclusion}
\label{sec:conclusion}

In this work, we have extended the thermodynamic geometric analysis of the
one-dimensional Blume--Capel model beyond the conventional Boltzmann--Gibbs
framework by incorporating the Tsallis nonextensive entropy. The resulting
thermodynamic metric and scalar curvature depend explicitly on the
nonextensive parameter $q$, allowing us to investigate how generalized
statistical weighting modifies the geometric structure of the thermodynamic
state space. By constructing the entropy-based thermodynamic metric as the negative Hessian of the generalized entropy in the parameter space $(\beta, J)$, we computed the associated scalar curvature $R(T)$ and analyzed its temperature dependence for both magnetically dominated ($D<J$) and crystal-field dominated ($D>J$) regimes. Although the model does not exhibit a true phase transition at finite temperature, the scalar curvature displays finite peaks that serve as clear geometric signatures of pseudo-critical crossovers. In the Boltzmann--Gibbs limit ($q=1$), these peaks reflect short-range attractive correlations and vanish smoothly at high temperature, consistent with the absence of genuine criticality in one dimension. The findings in the scalar curvature profiles are seen to be in line with that of the specific heat capacity curves where the curvature peaks are completely in line with the sudden fluctuations in the heat capacity profile for both the crystal field dominated and the magnetically dominated regime particularly for the $q>1$ case whereas we see no such behaviour in the $q<1$ case for both the regimes. \\

Our analysis demonstrates that the Tsallis deformation systematically alters the sign, magnitude, temperature location, and persistence of the scalar curvature peaks associated with pseudo-critical crossover behaviour. These results provide a direct information-geometric characterization of nonextensive effects in a spin-1 lattice model and establish a connection
between generalized entropy formalisms and thermodynamic geometry. For $q>1$, the suppression of rare configurations enhances the persistence of correlations and shifts the curvature peak, often maintaining nonvanishing curvature beyond the crossover region. In contrast, for $q<1$, the enhancement of rare states weakens collective magnetic behavior and can suppress or significantly reduce the curvature peak. Overall, our analysis demonstrates that generalized entropy induces measurable geometric deformations in the thermodynamic manifold, providing a clear information-geometric interpretation of non-extensive effects in low-dimensional spin-1 systems. \\

The present study has been restricted to a two-dimensional thermodynamic manifold parameterized by \((\beta,J)\), with the crystal-field anisotropy \(D\) treated as an external control parameter. An interesting extension of the present work would be to construct a higher-dimensional thermodynamic geometry by including \(D\) as an independent thermodynamic coordinate. Such a formulation would introduce additional metric components associated with anisotropy fluctuations and could provide further insight into the geometric role of the crystal field in spin-1 systems.\\

 One could also, in principle, construct an extended thermodynamic manifold using the coordinates \(x^{i}=(\beta,J,D)\), the resulting three-dimensional metric would contain additional mixed derivatives involving the crystal-field anisotropy. Since the principal objective of the present work is to investigate how the Tsallis nonextensive parameter modifies the thermodynamic geometry associated with thermal and exchange-interaction fluctuations, we restrict our analysis to the minimal two-dimensional manifold. A systematic investigation of the extended \((\beta,J,D)\) geometry is beyond the scope of the present work and will be considered in future studies.

\clearpage

\section{Acknowledgments}
\hspace{0.6cm}	The authors would like to thank Bidyut Hazarika and Mozib-Bin-Awal for the help they offered during the course of this work.

\bibliographystyle{unsrt} 
\bibliography{mybibfile}

\end{document}